\documentclass[iop,apj]{emulateapj}
\usepackage{amsmath}

\shorttitle{Noisy Field Line Twisting}
\shortauthors{Rappazzo, Velli, \& Einaudi}

\begin{document}

\title{Field lines twisting in a noisy corona: implications for 
energy storage and\\ release, and initiation of solar eruptions}

\author{A. F. Rappazzo$^{1,\dagger}$, M. Velli$^{2}$, and G.~Einaudi$^3$}
\affil{\vspace{.2em}
$^1$Bartol Research Institute, Department of Physics and Astronomy, University of Delaware, 
Delaware 19716, USA\\ 
\vspace{.2em}
$^2$Jet Propulsion Laboratory, California Institute of Technology, Pasadena, CA 91109, USA\\
\vspace{.2em}
$^3$Berkeley Research Associates, Inc., 6537 Mid Cities Avenue, Beltsville, MD 20705, USA}
\email{$^{\dagger}$rappazzo@udel.edu}

\begin{abstract}
We present simulations modeling closed regions of the solar corona threaded
by a strong magnetic field where localized photospheric vortical 
motions twist the coronal field lines.
The linear and nonlinear dynamics are 
investigated in the reduced magnetohydrodynamic regime in Cartesian geometry.
Initially the magnetic field lines get twisted
and the system becomes unstable to the internal kink mode, 
confirming and extending previous results.
As typical in this kind of investigations, where initial conditions
implement smooth fields and flux-tubes, we have neglected
fluctuations and the fields are laminar until the instability sets in.
But previous investigations indicate that fluctuations, excited 
by photospheric motions and coronal dynamics, are naturally
present at all scales in the coronal fields.
Thus, in order to understand 
the effect of a photospheric vortex on a more realistic corona, 
we continue the simulations after kink instability sets in,
when turbulent fluctuations have already developed in the corona.
In the nonlinear stage the system never returns to the simple initial 
state with ordered twisted field lines, and kink instability does not occur
again. Nevertheless field lines get twisted, but in a disordered way, and
energy accumulates at large scales through an inverse cascade.
This energy can subsequently be released in micro-flares or larger flares,
when interaction with neighboring structures occurs or via other mechanisms.
The impact on coronal dynamics and CMEs initiation is discussed.
\end{abstract}

\keywords{magnetohydrodynamics (MHD) --- Sun: corona --- Sun: coronal mass ejections (CMEs) --- 
Sun: magnetic topology --- turbulence}

\section{Introduction}

Photospheric convection and the coronal magnetic field play a key role in heating
the solar corona. For the magnetically confined (closed) regions of the  corona it 
has long been suggested that small heating events, dubbed \emph{nanoflares}, 
continuously deposit energy at the small scales and can contribute a substantial 
fraction of the total heating \citep{park72,park88,park94}.

The slow photospheric
motions (with a timescale $\tau_c \sim$ 8 minutes, magnitude $v_c \sim 1~km\, s^{-1}$, and correlation
scale  $\ell_c \sim 10^3\, km$) transfer energy from the photosphere into the corona
shuffling the footpoints of the magnetic field lines. The work done by the denser photospheric
plasma on the magnetic field lines footpoints injects energy into the corona,
mostly as magnetic energy.

The perturbations generated 
at the photospheric level propagate along the loop at the Alfv\'en speed.
In a coronal loop the Alfv\'en velocity associated to the strong axial magnetic field $B_0$ is
$v_A = B_0/\sqrt{4\pi\rho_0} \sim 2\times 10^3\, km\, s^{-1}$  ($\rho_0$ is the average mass density),
and considering a typical loop length $L \sim 4 \times 10^4\, km$ we obtain for
the Alfv\'en crossing time $\tau_A = L/v_A \sim 20\, s$. The crossing time is therefore 
significantly smaller than the photospheric timescale: $\tau_A << \tau_c$.
Furthermore as magnetic pressure largely exceeds plasma pressure the plasma $\beta$ is 
small ($<< 1$).
Because of the \emph{fast Alfv\'en timescale} and \emph{low} $beta$ the dynamics of the
magnetically confined solar corona are generally approximated as a \emph{quasi-static
evolution through a sequence of equilibria}, with instabilities leading the system
from an equilibrium to the next one through relaxation \citep{tay74, tay86, hp84}.
Within this framework many works have studied the relaxation dynamics in detail 
\citep[e.g.,][]{yhw10, pon11, ws11}.

On the other hand it has been shown that in some instances the validity of that 
picture is not valid, e.g., in reduced magnetohydrodynamic (MHD) simulations 
of the Parker model for coronal heating \citep{rved07,rved08,rve10}.
In fact that approximation is attained neglecting the velocity and plasma pressure in 
the MHD equations whose solution is then bound to be a static force-free equilibrium.
But the self-consistent evolution of the plasma pressure and velocity, although \emph{small} 
compared with the dominant axial magnetic field $B_0$ (and therefore very close in 
first approximation to the equilibrium solution of a uniform axial magnetic field), 
does not bind the system to force-free equilibria and allows the development of 
turbulent nonlinear dynamics with formation of field-aligned current sheets
where a significant heating occurs.

Most numerical simulations of a simple model coronal loop in Cartesian geometry threaded
by a strong magnetic field shuffled at the top and bottom plates by photospheric motions
have used as boundary velocity an incompressible field with all wavenumbers 
of order $\sim 4$ excited 
\citep{evpp96,dg97,gve98,dg99,ev99,dgm03,rved07,rved08,rv11}:
in real space this corresponds to distorted vortices with length-scale $\sim 10^3\, km$ 
one next to the other filling the whole photospheric plane \citep[][Figures~1 and 2]{rved08}. 
This configuration does not give rise to instabilities. The system transitions
smoothly from the linear to the nonlinear stage where integrated physical quantities like
energies and dissipation fluctuate around a mean value in what is best described as a 
statistically steady state and the energy deposited at the small scales is approximately
in the \emph{nanoflare} range for the numerous heating events.

This disordered vortical forcing mimics a uniform and homogeneous convection and 
the resulting coronal dynamics give rise to a basal background coronal heating 
within the lower limit of the observational constraint.

With this kind of forcing the system is not able to accumulate a significant amount
of magnetic energy to be subsequently released in more substantial heating events 
like \emph{microflares} and \emph{flares}. It is therefore pivotal to implement different
kinds of photospheric forcings to understand the role of convective motions in coronal 
\emph{heating} and the physical mechanisms and conditions for a significant \emph{storage} 
of magnetic energy and its \emph{release}.

Shearing or twisting the field lines might appear as the most straightforward way 
to make the system accumulate energy.
To this end, in recent work \citep{rve10} we have implemented a 1D 
velocity forcing with a sinusoidal \emph{shear flow} at the boundary, 
with $v_y(x) \propto sin(2\pi k x/\ell)$,  wavenumber $k=4$, spanning the 
whole photospheric plane ($\ell$ is the cross-section length).
Initially the magnetic field that develops in the coronal loop is a simple map 
of the photospheric velocity field, i.e.,
$b_y(x) \propto (t/\tau_A) sin(2\pi k x/\ell)$, with its intensity growing linearly in time. 
A sheared magnetic field is known to be subject to tearing instability \citep{fkr63}, 
in fact magnetic energy accumulates until a tearing instability sets in, 
magnetic energy is released and the system transitions to the nonlinear stage. 

On the other hand continuing the simulation we found that, 
once the system has become fully nonlinear the dynamics are 
fundamentally different: \emph{the magnetic shear is not recreated}. 
Once fluctuations are present, the orthogonal
magnetic field  ($b_x$ and $b_y$) is organized in
magnetic islands with $X$ and $O$ points, the nonlinear terms do not 
vanish any more and energy can be transported efficiently from large to small scales 
where it is dissipated. 

In the nonlinear stage the dynamics are very 
similar whether the forcing velocity is a shear flow or made of disordered 
vortices, and magnetic energy is not stored efficiently, therefore 
larger releases of energy are not possible.
A shear photospheric flow can give rise to a sheared coronal magnetic 
field only in the unlikely condition that no relevant perturbations are 
present in the corona, or for very strong shear flows with velocity 
higher than the typical photospheric velocity of $1\ km\, s^{-1}$, when the 
perturbations naturally present in the corona can be neglected.

All photospheric forcings described so far fill the entire photospheric plane. 
Spatially localized velocity fields, like a single vortex that does not fill the entire plane, 
might be able to induce a higher storage of magnetic energy in the corona.

A vortex twists the magnetic field lines, and the resulting 
helical magnetic structure is 
kink unstable and widely used to model coronal loops 
\citep{Baty:1996p5040, Velli:1997p4498, Lionello:1998p4478, Browning:2008p4608, Hood:2009p4102}.
In these studies a twisted magnetic field lines structure is used as initial condition,
but this is assumed to have been induced by photospheric motions shuffling their 
footpoints, while the field lines are actually line-tied to a \emph{motionless photosphere}.

To understand the dynamics of this photospherically driven system we have 
performed  numerical simulations applying localized vortices at 
the photospheric boundary. \cite{msvh90, Gerrard:2002p5041} 
performed boundary forced simulations, but they stop just after
kink instability sets in. In this paper we continue the simulations
for longer times. This allows us to understand the dynamics of the
system both when initially only an axial uniform magnetic field
is present and a smooth ordered flux-tube with twisted field lines
gets formed and kink instability sets in, and the later dynamics
when the localized boundary vortex twists the footpoints
of a disordered magnetic field where magnetic fluctuations
and small scales are already present and field lines are no longer smooth.

Furthermore, highly twisted magnetic structures, such as \emph{flux
ropes}, are broadly used to model the initiation of coronal mass ejections
(CMEs, e.g., see \citet{low01,al09,chen11,tor11}, and references therein).
To advance our understanding of eruption initiation
it is therefore important to understand if and under which conditions
photospheric motions can self-consistently generate flux ropes \citep{alm99}, or 
if these structures can only be advected into the corona from 
sub-photospheric regions via emerging flux.

The paper is organized as follows. In \S~\ref{sec:gebc} we describe
the basic governing equations 
and boundary conditions, as well as the numerical code used to integrate them.
In \S~\ref{par3} we discuss the initial conditions for our simulations and briefly summarize
the linear stage dynamics more extensively detailed in \cite{rved08}.
The results of our numerical simulations are presented in \S~\ref{sec:ns}, while
the final section is devoted to our conclusions and discussion of the
impact of this work on coronal physics.

\section{Governing Equations} \label{sec:gebc}

We model a coronal loop as an axially elongated Cartesian box with an orthogonal
cross section  of size $\ell$  and an axial length $L$  embedded in an homogeneous 
and uniform axial magnetic field $\mathbf{B_0} = B_0\, \mathbf{\hat{e}_z}$ aligned along
the $z$-direction. Any curvature effect is neglected.

The dynamics are integrated with the equations of RMHD \citep{kp74,str76,mon82},
well suited for a plasma embedded in a strong 
axial magnetic field. In dimensionless form they are given by:
{\setlength\arraycolsep{-10pt}
\begin{eqnarray}
&&\frac{\partial \mathbf{u_{_\perp}}}{\partial t}  + 
\left(  \mathbf{u_{_\perp}} \cdot \nabla_{_{\!\perp}} \right) \mathbf{u_{_\perp}} = 
- \nabla_{_{\!\perp}} \left( p + \frac{ \mathbf{b_{_\perp}}^2 }{2} \right)
\nonumber \\
&& \qquad + \left(  \mathbf{b_{_\perp}} \cdot \nabla_{_{\!\perp}} \right) \mathbf{b_{_\perp}}
+ c_{A}\, \frac{\partial \mathbf{b_{_\perp}}}{\partial z}
+ \frac{(-1)^{n+1}}{Re_n} \nabla_{_{\!\perp}}^{2n}\, \mathbf{u_{_\perp}}, 
\label{eq:eq1} \\
&&\frac{\partial \mathbf{b_{_\perp}}}{\partial t}  + 
\left(  \mathbf{u_{_\perp}} \cdot \nabla_{_{\!\perp}} \right) \mathbf{b_{_\perp}} = 
\left(  \mathbf{b_{_\perp}} \cdot \nabla_{_{\!\perp}} \right) \mathbf{u_{_\perp}} 
+ c_{A}\, \frac{\partial \mathbf{u_{_\perp}}}{\partial z}
\nonumber \\
&& \qquad \qquad \qquad \qquad \qquad \qquad \qquad 
+ \frac{(-1)^{n+1}}{Re_n} \nabla_{_{\!\perp}}^{2n}\, \mathbf{b_{_\perp}}, 
\label{eq:eq2} \\
&& \nabla_{_{\!\perp}} \cdot \mathbf{u_{_\perp}} = 0, \qquad 
 \nabla_{_{\!\perp}} \cdot \mathbf{b_{_\perp}} = 0,
\label{eq:eq3} 
\end{eqnarray}
}where $\mathbf{u_{_\perp}}$ and  $\mathbf{b_{_\perp}}$ are the velocity and magnetic
fields components orthogonal to the axial field and $p$ is the plasma pressure.
The gradient operator has components only in the perpendicular $x$-$y$ planes
\begin{equation}
\nabla_{_{\!\perp}} = \mathbf{\hat{e}}_x \frac{\partial}{\partial x} + 
\mathbf{\hat{e}}_y \frac{\partial}{\partial y}
\end{equation}
while the linear term $\propto \partial_z$ couples the planes along
the axial direction through  a wave-like propagation at the Alfv\'en 
speed $c_A$.
Incompressibility in RMHD equations follows from the large value
of the axial magnetic fields  \citep{str76} and they remain valid
also for low $\beta$ systems \citep{zm92,bns98} such as the corona.
Furthermore \cite{derv12} have recently performed fully compressible
simulations of a similar Cartesian coronal loop model, showing that the inclusion
of thermal conductivity and radiative losses, that transport away
the heat produced by the small scale energy dissipation,
keep the dynamics in the RMHD regime. 

To render the equations nondimensional, we have first expressed
the magnetic field as an Alfv\'en velocity [$b \rightarrow b/\sqrt{4\pi \rho_0}$],
where $\rho_0$ is the density supposed homogeneous and constant,
and then all velocities have been normalized to the velocity
$u^{\ast} = 1\ km\, s^{-1}$,  the order of magnitude of photospheric 
convective motions.

Lengths and times are expressed in units of the perpendicular length 
of the computational box $\ell^{\ast} = \ell$ and its related crossing time
$t^{\ast} = \ell^{\ast}/u^{\ast}$.
As a result, the linear terms $\propto \partial_z$ are multiplied by the
dimensionless Alfv\'en velocity $c_A = v_A/u^{\ast}$, where 
$v_A = B_0/\sqrt{4\pi \rho_0}$ is the Alfv\'en velocity associated with the axial magnetic field.
We use a computational box with an aspect ratio of $10$, which spans
\begin{equation}
0 \le x, y \le 1, \qquad 0 \le z \le 10.
\end{equation}
Our forcing velocities have a linear scale of $\sim 1/4$
that corresponds to the convective scale of $\sim 1,000\ \textrm{km}$ 
in conventional units,  thus the box extends $(4,000\ km)^2 \times 40,000\ km$. 

\begin{table}
\begin{center}
\caption{Simulations summary\label{tbl}}
\begin{tabular}{lccccr}
\hline \hline\\[-5pt]
Run & $n_x \times n_y \times n_z$ &  $Re_{_4}$ & $t_{max}$\\[5pt]
\hline\\[-5pt]
A     &  512 x 512 x 208     & $1\times 10^{19}$ &$  1,900\, \tau_{_{\!A}}$ \\[2pt]
B     &  256 x 256 x 192     & $5\times 10^{16}$ &$ 11,000\, \tau_{_{\!A}}$ \\[5pt]
\hline
\end{tabular}
\tablecomments{The numerical grid resolution is $n_x \times n_y \times n_z$. 
The next columns indicate respectively the value of the 
hyperdiffusion coefficient $Re_{_4}$ and the simulation time span.}
\end{center}
\end{table}

The index $n$ in the diffusive terms (\ref{eq:eq1})-(\ref{eq:eq2})
is called \emph{dissipativity} and for $n > 1$ 
these correspond to so-called hyperdiffusion \citep[e.g.,][]{bis03}.
For $n=1$ standard diffusion ($Re_{_1} = Re$) is recovered and in this case
the kinetic and magnetic Reynolds numbers are given by:
\begin{equation}
Re = \frac{\rho_0\, \ell^{\ast} u^{\ast}}{\nu}, \qquad 
Re_{_m} = \frac{4\pi\, \ell^{\ast} u^{\ast}}{\eta c^2},
\end{equation}
where $c$ is the speed of light, and numerically they are given
the same value $Re=Re_{_m}$.

In the simulations presented in this paper we use hyperdiffusion with $n=4$.
Hyperdiffusion is used because the implemented boundary velocity forcings
and the magnetic flux tubes  induced  initially are localized to a small area of
the computational box, and the dynamics would be dramatically diffusive
with standard diffusion at a reasonable resolution
(see next section \S~\ref{par3} for a more detailed discussion).

Our parallel code RMH3D solves numerically 
Equations~(\ref{eq:eq1})-(\ref{eq:eq3}) 
written in terms of the potentials of the orthogonal
velocity and magnetic fields in Fourier space, i.e., we advance the Fourier components 
in the $x$- and $y$-directions of the scalar potentials. Along 
the $z$-direction, no Fourier transform is performed 
so that we can impose non-periodic boundary conditions (\S~\ref{par3}), 
and a central second-order finite-difference
scheme is used. In the $x$-$y$ plane, a Fourier pseudospectral method 
is implemented. Time is discretized with a third-order Runge-Kutta method.
For a more detailed description of the numerical code see \cite{rved07,rved08}.

\section{Boundary Conditions and Linear Stage Dynamics}  \label{par3}

Magnetic field lines are line-tied to the top and bottom plates ($z=0$ and $L$) 
that represent the photospheric surfaces. Here we impose,  as boundary condition,  
a velocity field that convects the footpoints of the magnetic field lines.
Along the $x$ and $y$ directions periodic boundary conditions are implemented.

All simulations  (see Table~\ref{tbl}) employ a circular vortex 
applied at the top plate $z=L$. The
velocity potential for this vortex is centered in the interval 
$\mathit{I} = \left[ 1/2 - 1/8, 1/2+1/8 \right]$ of linear extent $\ell_c = 1/4$ 
and vanishes outside:{\small 
\begin{eqnarray}
&&\varphi\! \left( x, y \right)  =  \frac{1}{2\pi \sqrt{3}}\,  
\sin^2\! \left[ 4\pi \left( x - \frac{1}{8} \right) \right] 
\sin^2\! \left[ 4\pi \left( y - \frac{1}{8} \right) \right]  \nonumber \\
&&\mathrm{for} \quad x, y \in \mathit{I}, \quad \mathrm{and} \quad
\varphi = 0 \quad  \mathrm{for} \quad x, y \notin \mathit{I}. \label{eq:f0}
\end{eqnarray}
}The velocity is linked to the potential by
$\mathbf{u}_{_\perp}  = \nabla \varphi \times \mathbf{\hat{e}}_z$
and its components are:{\small
\begin{eqnarray} 
&&u_x^L\! \left( x, y \right)  = +\frac{2}{\sqrt{3}}\,  
\sin^2\! \left[ 4\pi \left( x - \frac{1}{8} \right) \right] 
\sin\! \left[ 8\pi \left( y - \frac{1}{8} \right) \right]  \label{eq:f01}\\
&&u_y^L\! \left( x, y \right)  = -\frac{2}{\sqrt{3}}\,  
\sin\! \left[ 8\pi \left( x - \frac{1}{8} \right) \right]
\sin^2\! \left[ 4\pi \left( y - \frac{1}{8} \right) \right] \label{eq:f00}
\end{eqnarray}
}in the interval \textit{I} and vanish outside.
\begin{figure}
\begin{centering}
\includegraphics[scale=.57]{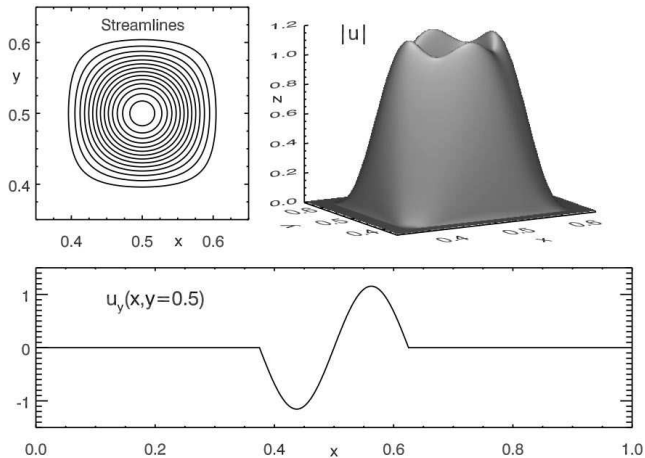}
\caption{Circular vortex employed as boundary velocity forcing
in the presented simulations.
\emph{Above:} streamlines and profile of its absolute value $|u|$.
\emph{Below:} plot of the velocity $y$-component as a function of
$x$ at $y=0.5$.
\label{fig:fig1}}
\end{centering}
\end{figure}
As shown in Figure~\ref{fig:fig1} Equations~(\ref{eq:f0})-(\ref{eq:f00}) describe a 
counter-clockwise vortex 
centered in the middle of the plane $z=L$ and has circular streamlines, with a slight
departure from a perfectly circular shape toward the edge of the interval \textit{I}.
Averaging over the surface \textit{I} the velocity rms is $\langle (u^L)^2 \rangle_I = 1/2$,
the same value of the boundary velocity fields used in our previous works
\citep{rved07,rved08,rve10,rv11}. 

In all simulations a vanishing velocity is imposed at the bottom plate $z=0$:
\begin{equation} 
\mathbf{u}^0 \left( x, y \right)  =  0. \label{eq:f1}
\end{equation}

At time $t=0$ we start our simulations with a uniform and homogeneous magnetic field
along the axial direction $\mathbf{B}_0 = B_0\, \mathbf{\hat{e}}_z$. 
The orthogonal component of the velocity and magnetic fields are zero
inside our computational box $\mathbf{u_{_\perp}}=\mathbf{b_{_\perp}}=0$,
while at the top and bottom planes the vortical velocity forcing is applied
and kept constant in time. 

We briefly summarize and specialize to the case considered in this paper
the linear stage analysis described in more detail
in \cite{rved08}. In general for an initial interval of time smaller than the 
nonlinear timescale $t < \tau_{nl}$,  nonlinear terms in Equations 
(\ref{eq:eq1})-(\ref{eq:eq3}) can be neglected and the equations linearized. 
For simplicity we will at first neglect also the diffusive terms and consider
their effect later in this section. 
The solution during the linear stage with a generic boundary 
velocity forcing $\mathbf{u}^L$, and $\mathbf{u}^0 = 0$, (respectively at the top
and bottom planes $z=L$ and $0$) is given by:{\setlength\arraycolsep{-10pt}
\begin{eqnarray}
&&\mathbf{b_{_\perp}} (x,y,z,t) = 
\frac{t}{\tau_A}\, \mathbf{u}^L, 
\label{eq:lin1} \\
&&\mathbf{u_{_\perp}} (x,y,z,t) = 
\frac{z}{L}\, \mathbf{u}^L.
\label{eq:lin2}
\end{eqnarray}
}where $\tau_A = L/v_A$ is the Alfv\'en crossing time along the axial direction $z$.
The magnetic field grows linearly in time, while the velocity field
is stationary and the order of magnitude of its rms is determined
by the boundary velocity profile.
Both are a \emph{mapping} of the boundary velocity field $\mathbf{u}^L$.

For a generic forcing the solution  (\ref{eq:lin1})-(\ref{eq:lin2}) 
is valid only during the linear stage, while for $t > \tau_{nl}$ when the fields 
are big enough the nonlinear terms cannot be neglected.
Nevertheless there is a singular subset of velocity forcing
patterns for which the generated coronal fields (\ref{eq:lin1})-(\ref{eq:lin2})
have a vanishing Lorentz force and the nonlinear terms vanish exactly.

This subset  of patterns is characterize by having the \emph{vorticity constant
along the streamlines} \citep{rved08}.
In this case \emph{magnetic energy grows} 
quadratically in time until some \emph{instability} eventually sets in.
Two kind of velocity patterns can be identified: a) 1D patterns with their streamlines
all parallel to each other, like a \emph{shear flow}, or b) a radial pattern with
circular streamlines, like a \emph{circular vortex}.

Since in the linear stage the coronal fields are a mapping of the boundary
velocity (\ref{eq:lin1})-(\ref{eq:lin2}), a shear flow induces a sheared magnetic
field subject to tearing instabilities \citep{rve10}, while the vortical flows
considered in this paper twist the field lines into helices subject to kink instabilities.
The vortex  (\ref{eq:f01})-(\ref{eq:f00}) is not perfectly circular as the streamlines
depart from an exact round shape toward the edge (Figure~\ref{fig:fig1}), but 
as we show in \S~\ref{sec:runa} field line tension adjusts
the induced coronal orthogonal field lines in a round shape.

So far we have neglected the diffusive terms in the RMHD Equations
(\ref{eq:eq1})-(\ref{eq:eq3}). In \S~\ref{sec:ns} we show that for
this kind of problem the use of hyperdiffusion is crucial, otherwise
the dynamics are dominated by diffusion. Overlooking 
this numerical fact can result in misleading conclusions
\citep{knda09,knda10}, upon which we will comment in \S~\ref{sec:ns}.

Here we want to understand the diffusive effects on the linear dynamics,
i.e.,  when nonlinear terms are negligible or artificially suppressed 
by a low numerical resolution.
We now consider the effect of standard diffusion 
(case $n=1$ in eqs.~(\ref{eq:eq1})-(\ref{eq:eq2})) on the solutions
(\ref{eq:lin1})-(\ref{eq:lin2}): these are the solutions of the linearized
equations obtained from (\ref{eq:eq1})-(\ref{eq:eq2}) retaining
also the diffusive terms.

In the \emph{linear regime}, as the magnetic field grows in time 
(\ref{eq:lin1}), the diffusive term 
($\nabla_{_{\!\perp}}^2\, \mathbf{b_{{_\perp}}} \! \propto \! \mathbf{b}_{_\perp}/\ell^2$)
becomes increasingly bigger until diffusion balances the magnetic field growth, 
and the system reaches a saturated equilibrium state. 
Including diffusion the magnetic field evolves as
\begin{equation}
\mathbf{b_{_{\!\perp}}} (x,y,z,t) = \mathbf{u}^L (x,y)
\frac{\tau_R}{\tau_A} \, 
\left[ 1 - \exp \left( - \frac{t}{\tau_R} \right) \right], \label{eq:diff1}
\end{equation}
i.e., for times smaller than the diffusive timescale $\tau_R$ Equation~(\ref{eq:lin1})
is recovered with the field growing linearly in time, while for times bigger
than $\tau_R$ the field asymptotes to its saturation value. 
The diffusive timescale associated with the Reynolds number
$Re$ is  $\tau_R = \ell_c^2\, Re / (2\pi)^2$ where $\ell_c$ is the
length-scale of the forcing pattern, that for the pattern (\ref{eq:f0})-(\ref{eq:f00})
is given by $\ell_c \sim \ell / 4$ where $\ell$ is the orthogonal computational box length.

The total magnetic energy $E_M$ and ohmic dissipation rate $J$ 
will then be given by
{\setlength\arraycolsep{-20pt}
\begin{eqnarray}
&&E_M = \frac{1}{2}\, \int_V\! \mathrm{d}^{^3}\hspace{-.4em} x \ \mathbf{b}_{_\perp}^2 = 
E_M^{sat} \, \left[ 1 - \exp \left( - \frac{t}{\tau_R} \right) \right]^2,
\label{eq:diff2} \\
&&J = \frac{1}{Re}\, \int_V\! \mathrm{d}^{^3}\hspace{-.4em}  x \ \mathbf{j}^2 = 
J^{sat} \, \left[ 1 - \exp \left( - \frac{t}{\tau_R} \right) \right]^2.
\label{eq:diff3}
\end{eqnarray}
}For times smaller than the diffusive timescale $\tau_R$ both quantities
grow \emph{quadratically} in time, while for $t \gtrsim 2\, \tau_R$ they 
asymptote to their saturation value $E_M^{sat}$ and $J^{sat}$:
\begin{equation}
E_M^{sat} = \frac{\ell_c^6 c_A^2 Re^2}{2L(2\pi)^4}\, \langle (u^L)^2 \rangle_I, \quad
J^{sat}  = \frac{2\, E_M^{sat}}{(4\pi)^2 \ell_c^2  Re}.
\label{eq:diff4}
\end{equation}
Magnetic energy saturates to a value proportional to the square of both the Reynolds number
and the Alfv\'en velocity, while the heating rate saturates to a value that is proportional to the 
Reynolds number and the square of the axial Alfv\'en velocity.

Even though  we use  grids  with $\sim 512^2$ points in the x-y plane,
the timescales associated with ordinary diffusion are small enough to affect 
the large-scale dynamics, inhibiting the development of instabilities and nonlinearity.
The diffusive time $\tau_{_n}$ at the scale $\lambda$ 
associated with the dissipative terms used in 
Equations~(\ref{eq:eq1})-(\ref{eq:eq2}) is given by
\begin{equation}
\tau_{_n} \sim Re_{_n}\, \lambda^{2n}.
\end{equation}
For $n=1$ the diffusive time decreases relatively slowly toward smaller scales, 
while for $n=4$ it decreases far more rapidly. As a result for $n=4$ we have 
longer diffusive timescales at large spatial scales and diffusive timescales 
similar to the case with $n=1$ at the resolution scale. 
Numerically we require the diffusion time at the 
resolution scale $\lambda_{min} = 1/N$, where N is the number of grid points, to be of 
the same order of magnitude for both normal and hyper-diffusion, i.e.,
\begin{equation}
\frac{Re_{_1}}{N^2} \sim \frac{Re_{_n}}{N^{2n}} \quad
\longrightarrow \quad
Re_{_n} \sim Re_{_1}\, N^{2(n-1)}.
\end{equation}
Then for a numerical grid with $N=512$ points that requires
a Reynolds number $Re_{_1} = 800$ with ordinary diffusion we can 
implement $Re_{_{4}} \sim 10^{19}$ (table~\ref{tbl}), removing diffusive effects at the 
large scales and allowing, if present, the development of kink instabilities
and nonlinear dynamics.

\begin{figure}
\begin{centering}
\includegraphics[scale=.57]{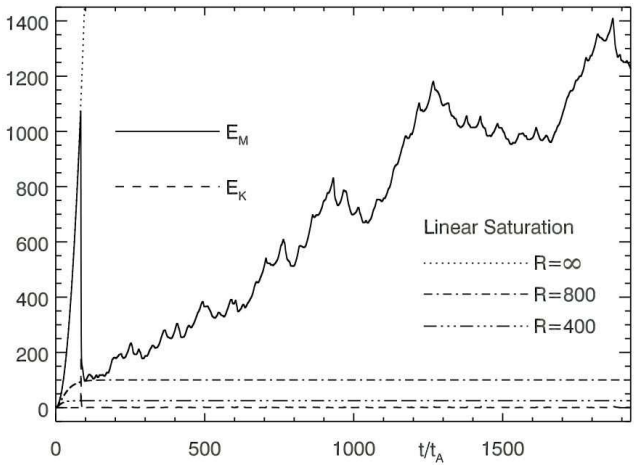}
\caption{\emph{Run~A:} Magnetic ($E_M$) and kinetic ($E_K$) energies
as a function of time ($\tau_A = L/v_A$ is the axial Alfv\'en crossing time).
The dashed curves show the time evolution of magnetic energy
if the system were unperturbed [eq.~(\ref{eq:diff2})], or 
nonlinearity suppressed, e.g., by numerical diffusion (see \S\ref{par3}).
The case $R=\infty$ corresponds to the linear case with no diffusion,
attained (at the large scales) with the implementation of hyperdiffusion.
\label{fig:fig2}}
\end{centering}
\end{figure}

\section{Numerical Simulations} \label{sec:ns}
 
In this section we present the results of the numerical simulations summarized in 
Table~\ref{tbl}. Simulations A and B have the same parameters, but
simulation B employs a lower resolution to achieve a very long duration.
In all simulations the vortical velocity pattern (\ref{eq:f0})-(\ref{eq:f00}) is applied
at the top plate $z=L$, and a vanishing velocity at the bottom plate 
$z=0$. Initially no perpendicular magnetic or velocity field is 
present inside the computational box 
$\mathbf{b}_{_{\!\perp}} = \mathbf{u}_{_{\!\perp}} = 0$, and the system is threaded
only by the constant and uniform field $\mathbf{B}_0 = B_0\, \mathbf{\hat{e}}_z$.
The computational box has an aspect ratio of $10$,
with $\ell = 1$ and $L=10$.

\subsection{Run~A} \label{sec:runa}

We present here the results of run~A, a simulation performed with a numerical grid
of $512 \times 512 \times 208$ points, and hyperdiffusion coefficient $Re_4 = 10^{19}$
with diffusivity $n=4$.
The Alfv\'en velocity is $v_A = 200\, \textrm{km}\, \textrm{s}^{-1}$, corresponding to a nondimensional
ratio $c_A = v_A/u^{\ast} = 200$. The total duration is  $\sim 1,900$ axial Alfv\'en crossing
times $\tau_A = L/v_A$. 

\begin{figure}
\begin{centering}
\includegraphics[scale=.57]{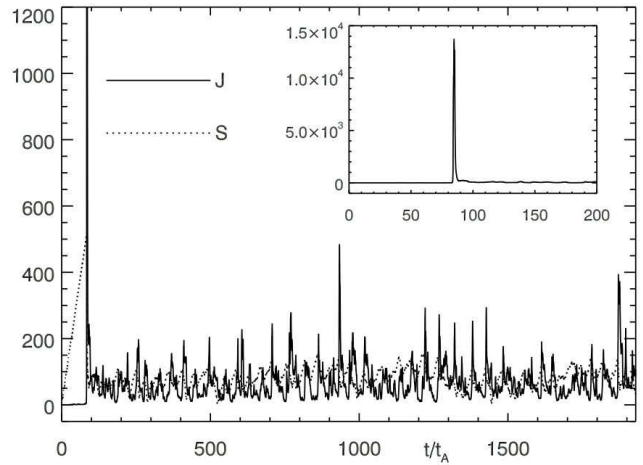}
\caption{\emph{Run~A:} Ohmic ($J$) dissipation rate and
the integrated Poynting flux $S$ (the injected power) versus 
time. Viscous dissipation is negligible respect to the
ohmic contribution. Inset shows the ohmic dissipative peak
corresponding to the development of kink instability. 
\label{fig:fig3}}
\end{centering}
\end{figure}

\begin{figure*}
\begin{centering}
\includegraphics[scale=.50]{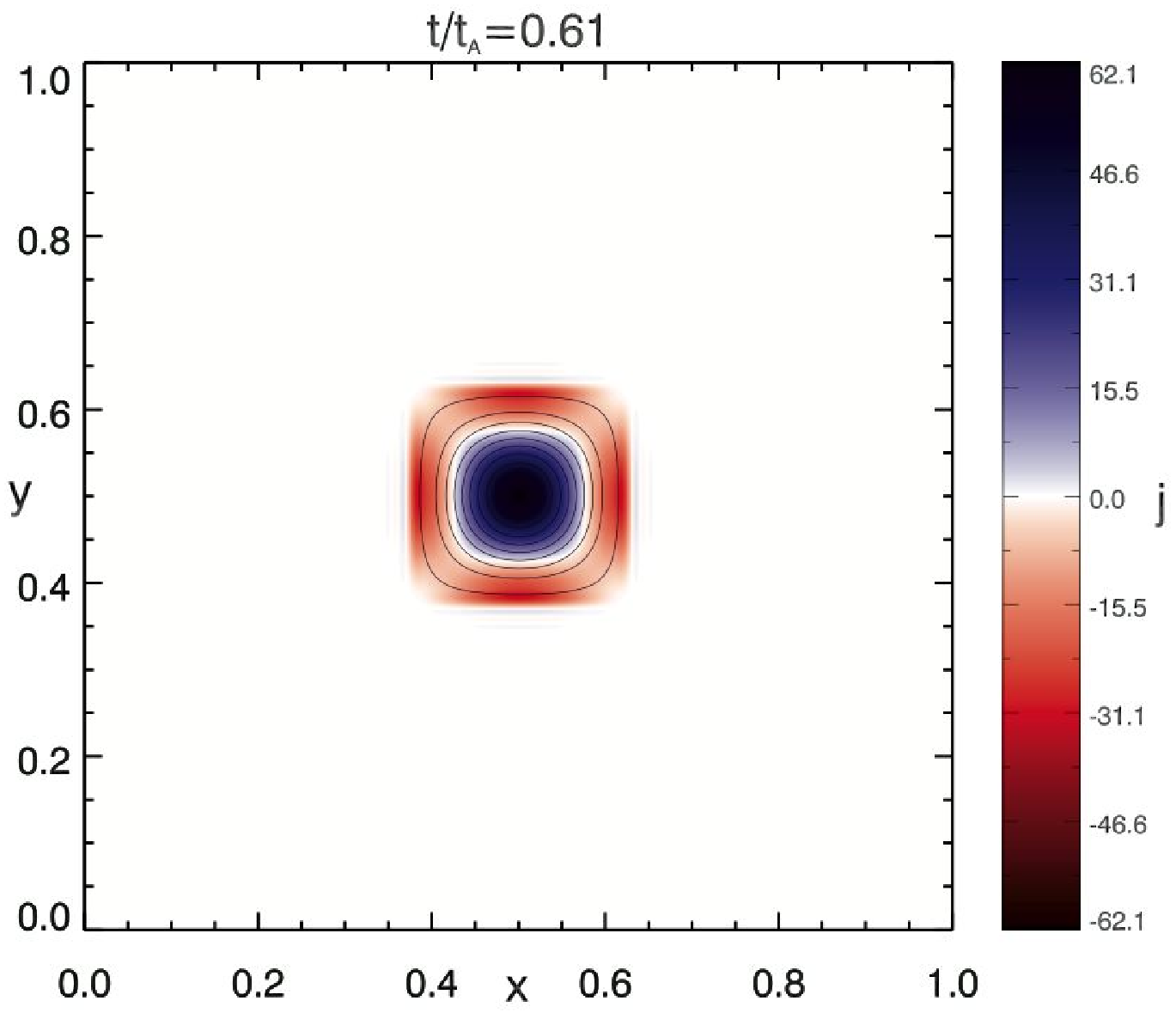}
\includegraphics[scale=.50]{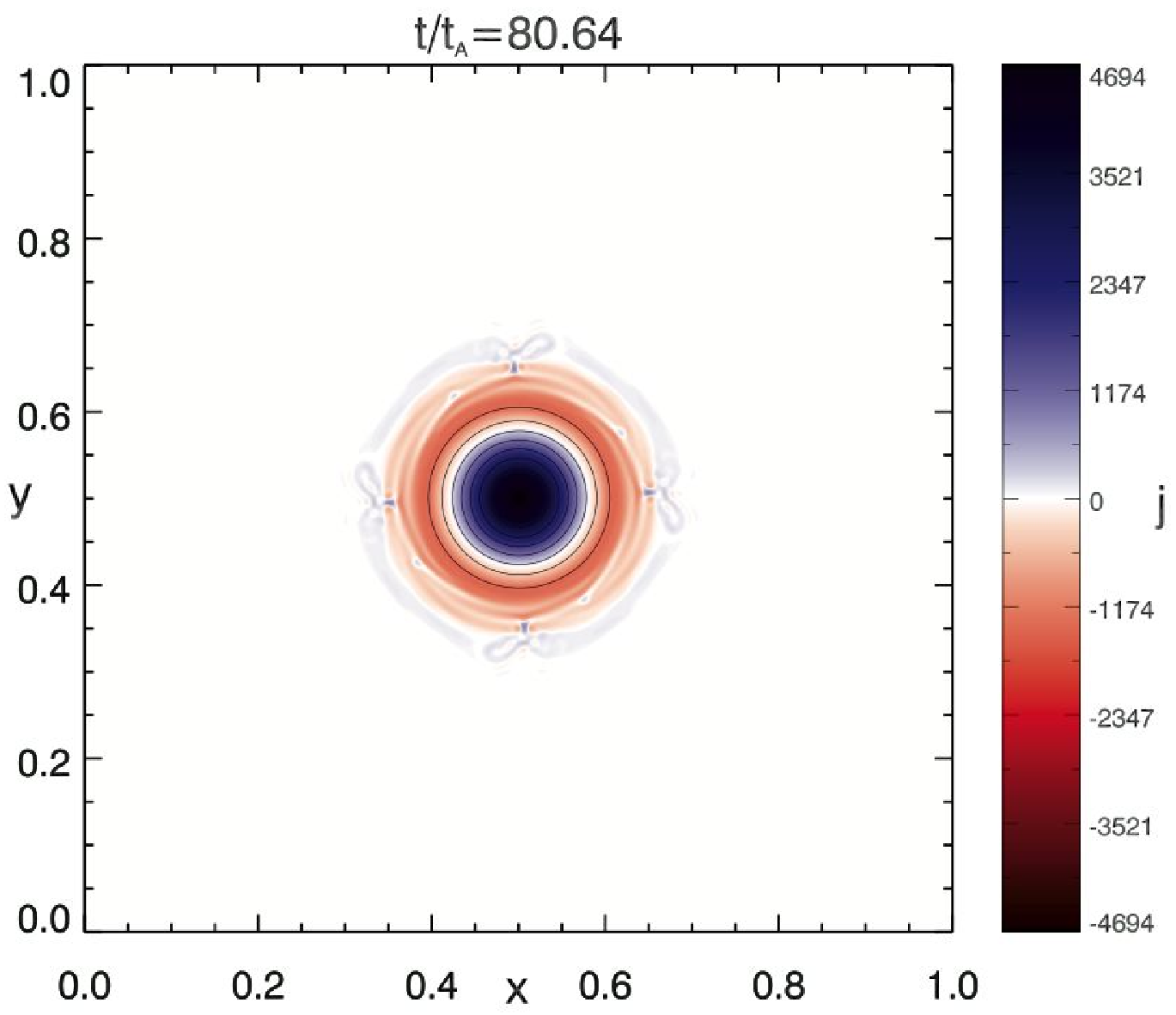}\\[.3em]
\includegraphics[scale=.50]{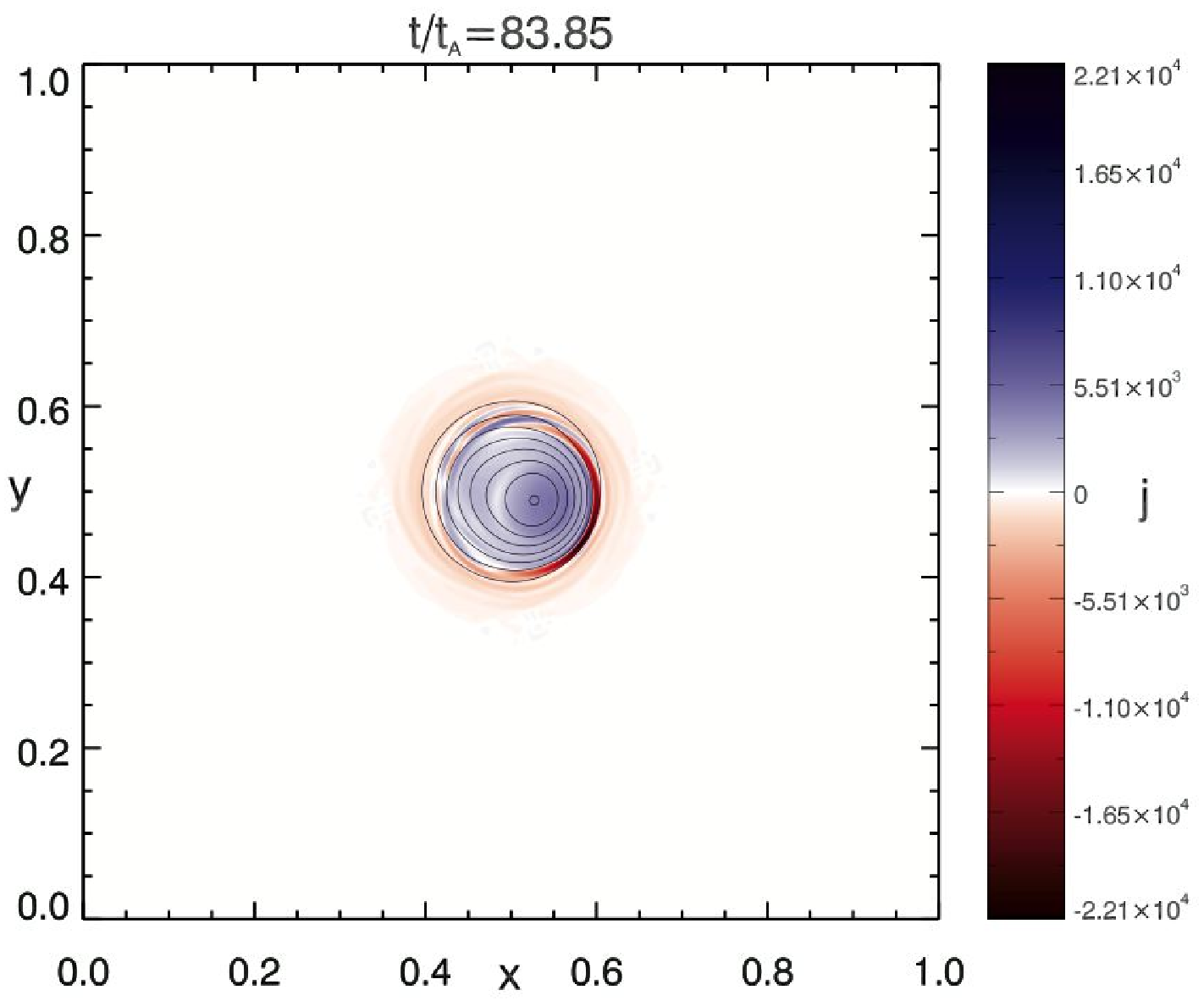}
\includegraphics[scale=.50]{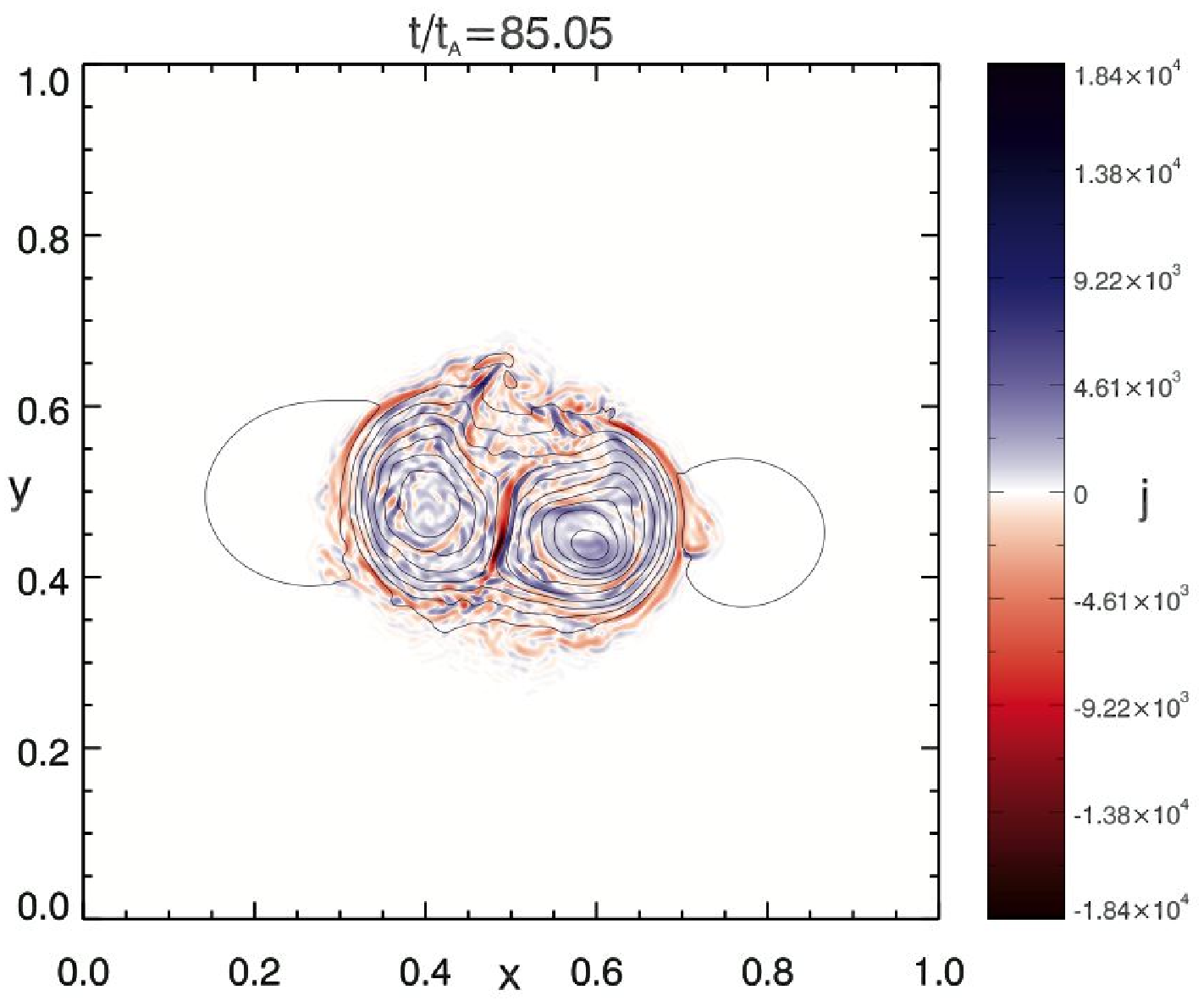}\\[.3em]
\includegraphics[scale=.50]{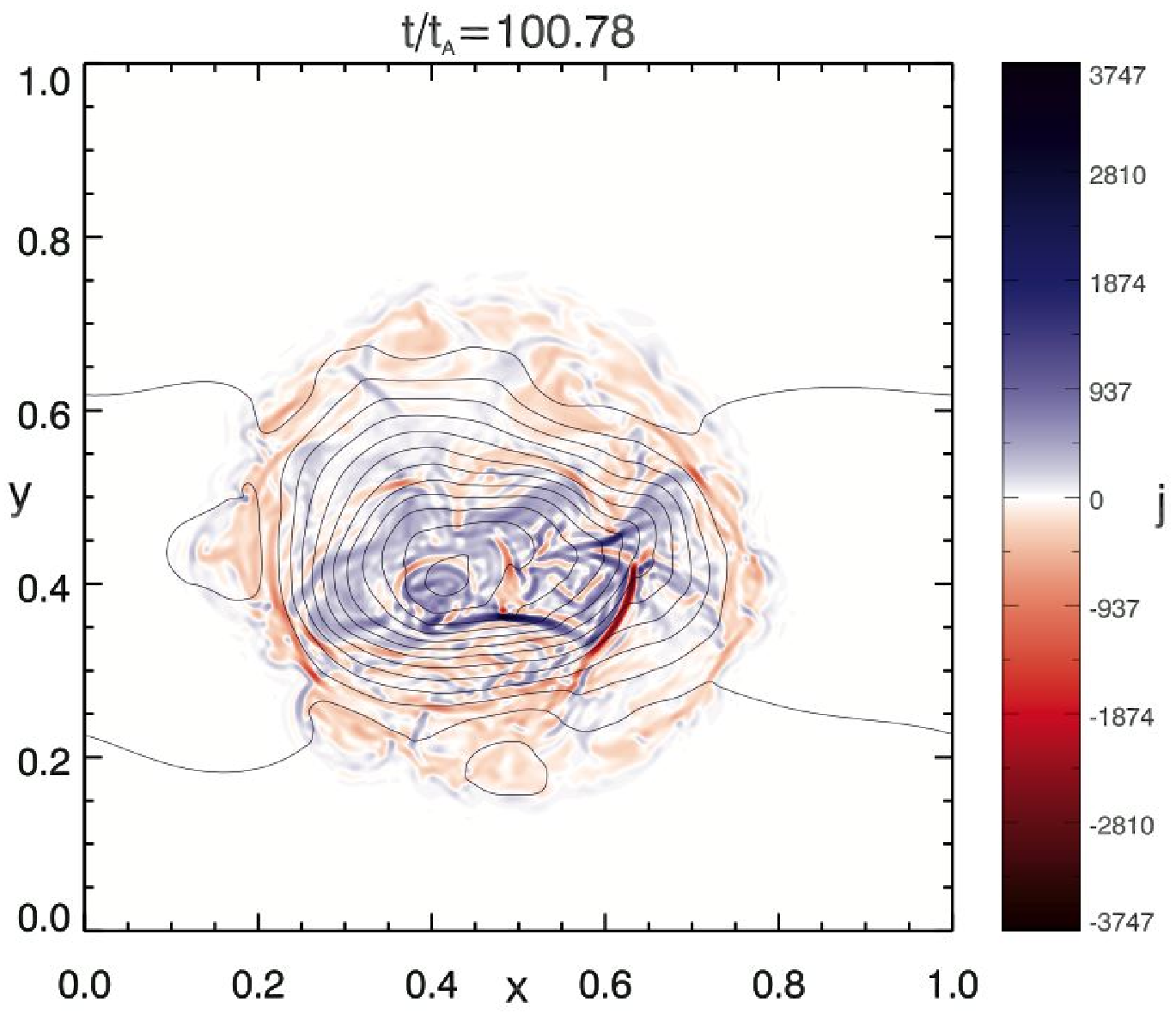}
\includegraphics[scale=.50]{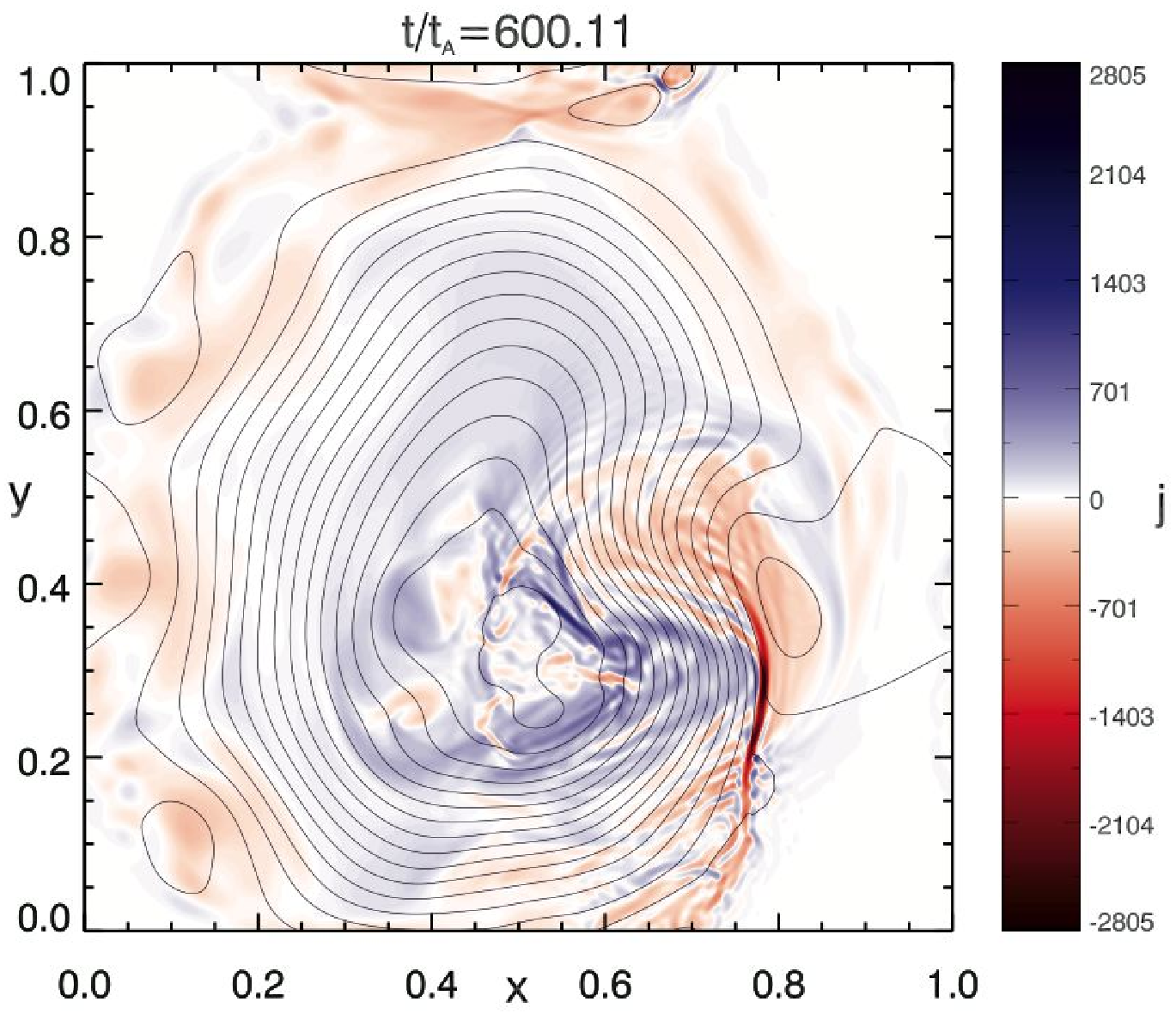}
\caption{\emph{Run~A}: Axial component of the current $j$ (in color) and field lines 
of the orthogonal magnetic field in the midplane ($z=5$) at selected times covering 
the linear and nonlinear regimes up to $t\sim 600\, \tau_A$.
At the beginning of the linear stage ($t = 0.61\, \tau_A$)
the orthogonal magnetic field is a mapping of the boundary vortex
[see linear analysis, Equation~(\ref{eq:lin1})].
Still in the linear stage but at later times ($t = 80.64\, \tau_A$)
the field line tension straightens out in a circular shape
the vortex mapping. An internal kink mode develops ($t \sim 83.85\, \tau_A$)
and the instability transitions the system to the nonlinear stage.
In the fully nonlinear stage the field lines are still circular, but in a disordered
way, exhibit a broad range of scales, including current sheets, and steadily 
occupy a larger fraction of the computational box.
\label{fig:fig4}}
\end{centering}
\end{figure*}

Figures~\ref{fig:fig2}-\ref{fig:fig3} show the temporal evolution of  the total magnetic 
and kinetic energies
\begin{equation}
E_M = \frac{1}{2} \int \! \mathrm{d}V\, \mathbf{b}_{_{\!\perp}}^2, \qquad
E_K = \frac{1}{2} \int \! \mathrm{d}V\, \mathbf{u}_{_{\!\perp}}^2, 
\end{equation}
the total ohmic dissipation rate
\begin{equation}
J             = \frac{1}{Re} \int \! \mathrm{d}V\, \mathbf{j}^2, 
\end{equation}
and $S$, the power injected from the boundary by the work done by
convective motions on the field lines' footpoints (see Equation~(\ref{eq:pflux})),
along with some saturation  curves for magnetic energy~(\ref{eq:diff2}).
Additionally Figure~\ref{fig:fig4} shows  snapshots of the magnetic field lines 
of the orthogonal component $\mathbf{b_{_\perp}}$ and
electric current $j=j_z$, the leading order component in RMHD
ordering \citep{str76}, at selected times in the mid-plane $z=5$.

The circular vortical velocity field~(\ref{eq:f01})-(\ref{eq:f00}) applied at the top
boundary ($z=10$) initially induces velocity and magnetic fields in the computational 
box that follow the linear behavior given by Equations~(\ref{eq:lin1})-(\ref{eq:lin2}),
i.e., they are a mapping of the velocity at the boundary with the magnetic field
increasing linearly in time (Figure~\ref{fig:fig4}, times $t = 0.61\, \tau_A$ 
and $80.64\, \tau_A$). 
In the linear stage ($t \lesssim 83\, \tau_A$) magnetic energy is well-fitted 
(Figure~\ref{fig:fig2}) by  the linear  curve~(\ref{eq:diff2}) in the limit $Re \rightarrow \infty$,
i.e., in the absence of diffusion (indeed in this limit the curve can be obtained directly from 
the linear Equation~(\ref{eq:lin1})). This is because we are using hyperdiffusion
that effectively gets rid of diffusion at the large scales.

Two other magnetic energy linear diffusive saturation curves are 
drawn for $Re = 800$ and $400$, typical
Reynolds numbers  used in our previous simulations with standard diffusion $n=1$ and
orthogonal grids with respectively $512^2$ and $256^2$ grid points \citep[see, e.g.,][]{rved08}.
Their saturation level is very low compared to the magnetic energy values
when kink instability develops ($t \sim 83\, \tau_A$) and in the following
nonlinear stage. This is because the vortex and induced magnetic field
occupy only a limited volume elongated along z at the center of the x-y plane:
at these scales diffusion dominates with these resolutions using standard diffusion.

\begin{figure*}
\begin{centering}
\includegraphics[scale=.35]{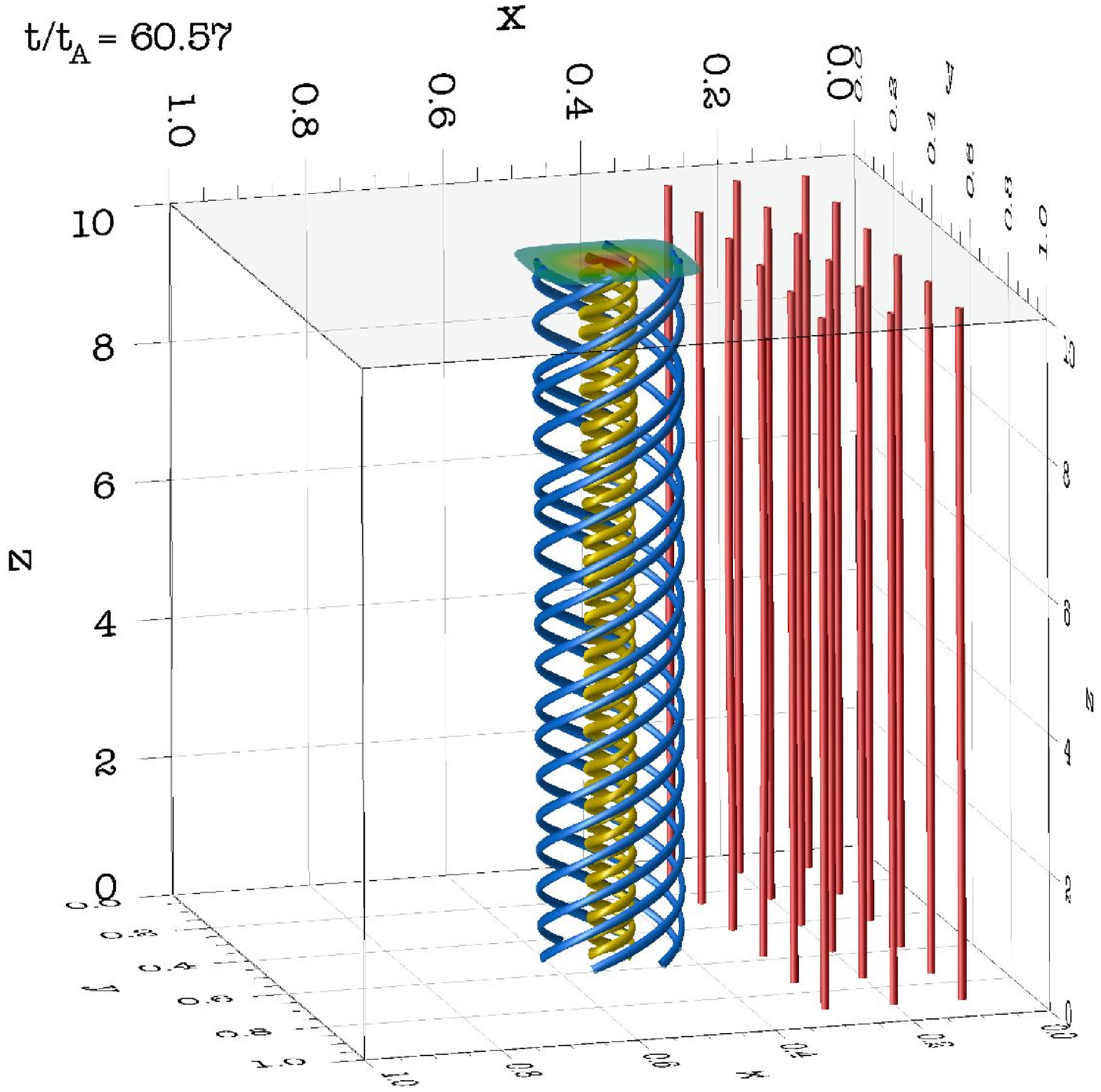}\hspace{1em}
\includegraphics[scale=.35]{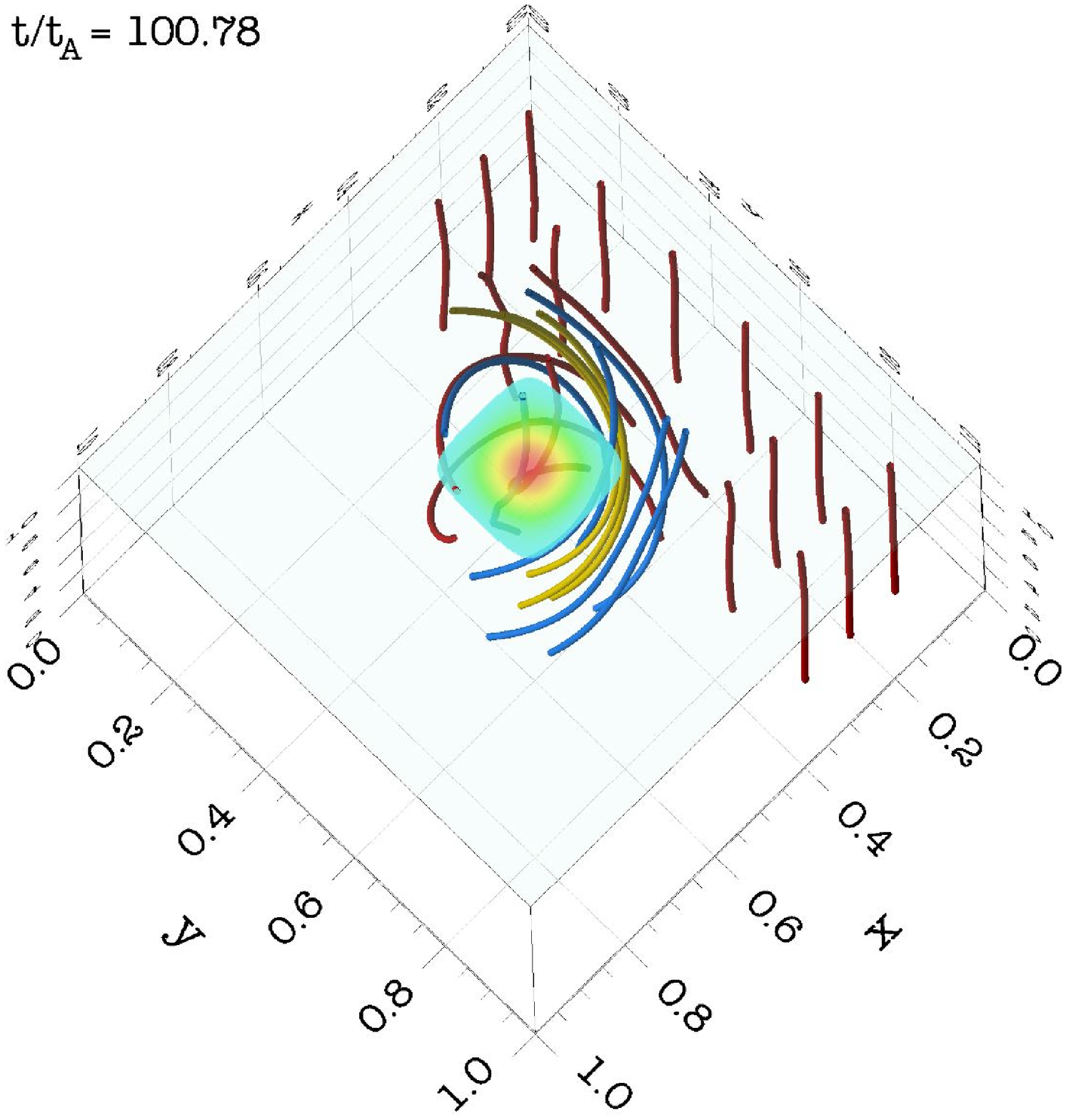}\\[.5em]
\includegraphics[scale=.35]{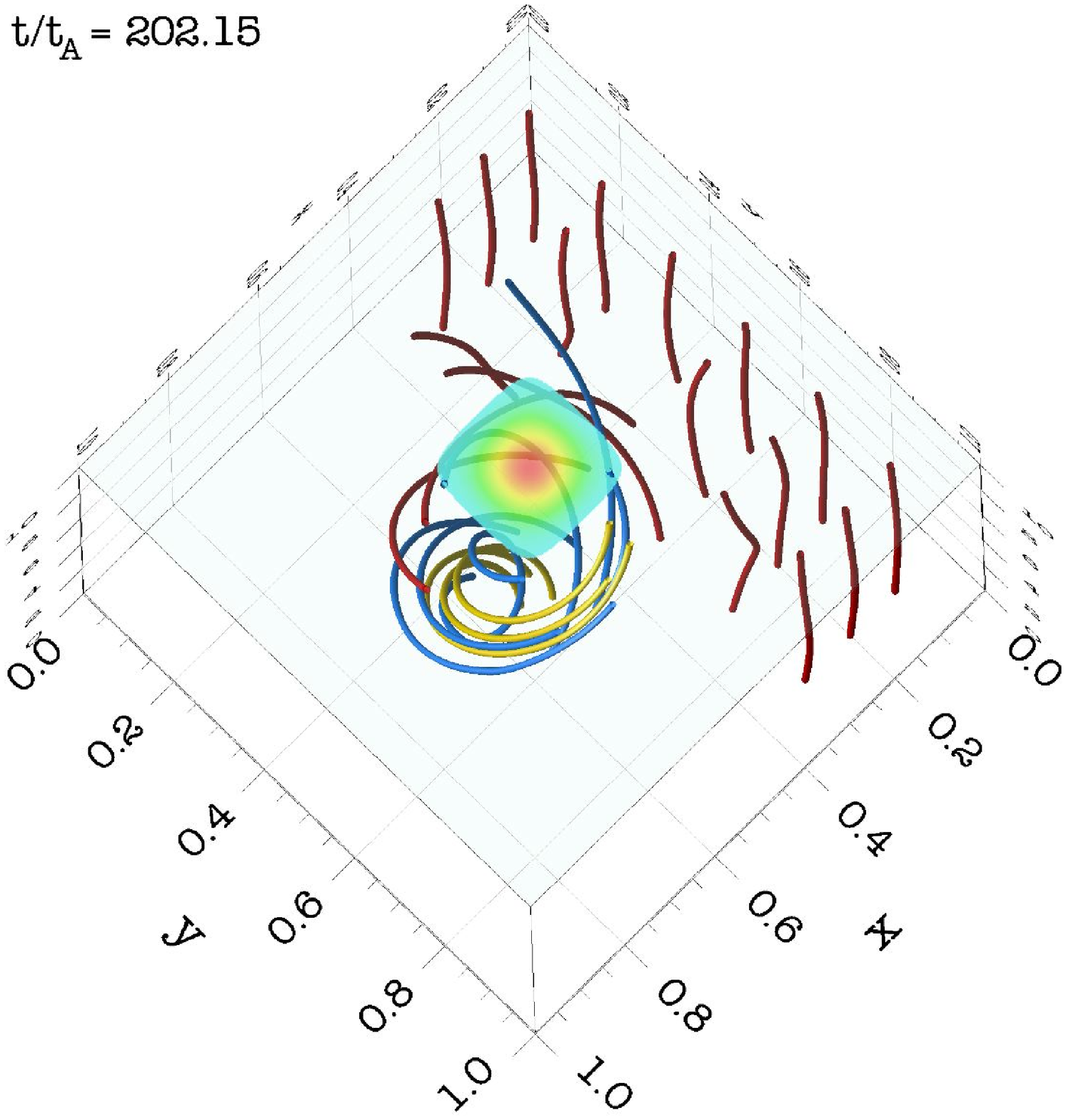}\hspace{1em}
\includegraphics[scale=.35]{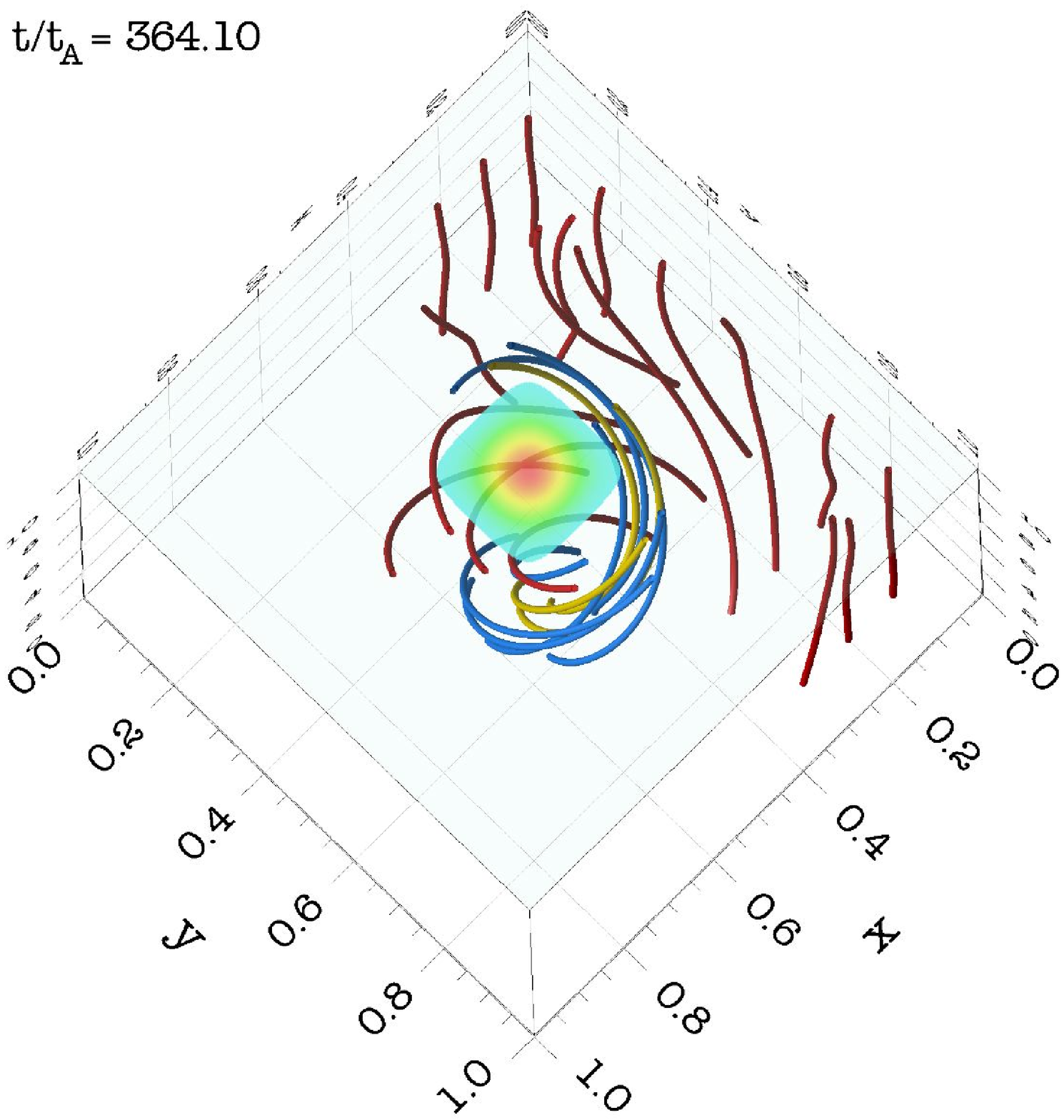}\\[.5em]
\includegraphics[scale=.35]{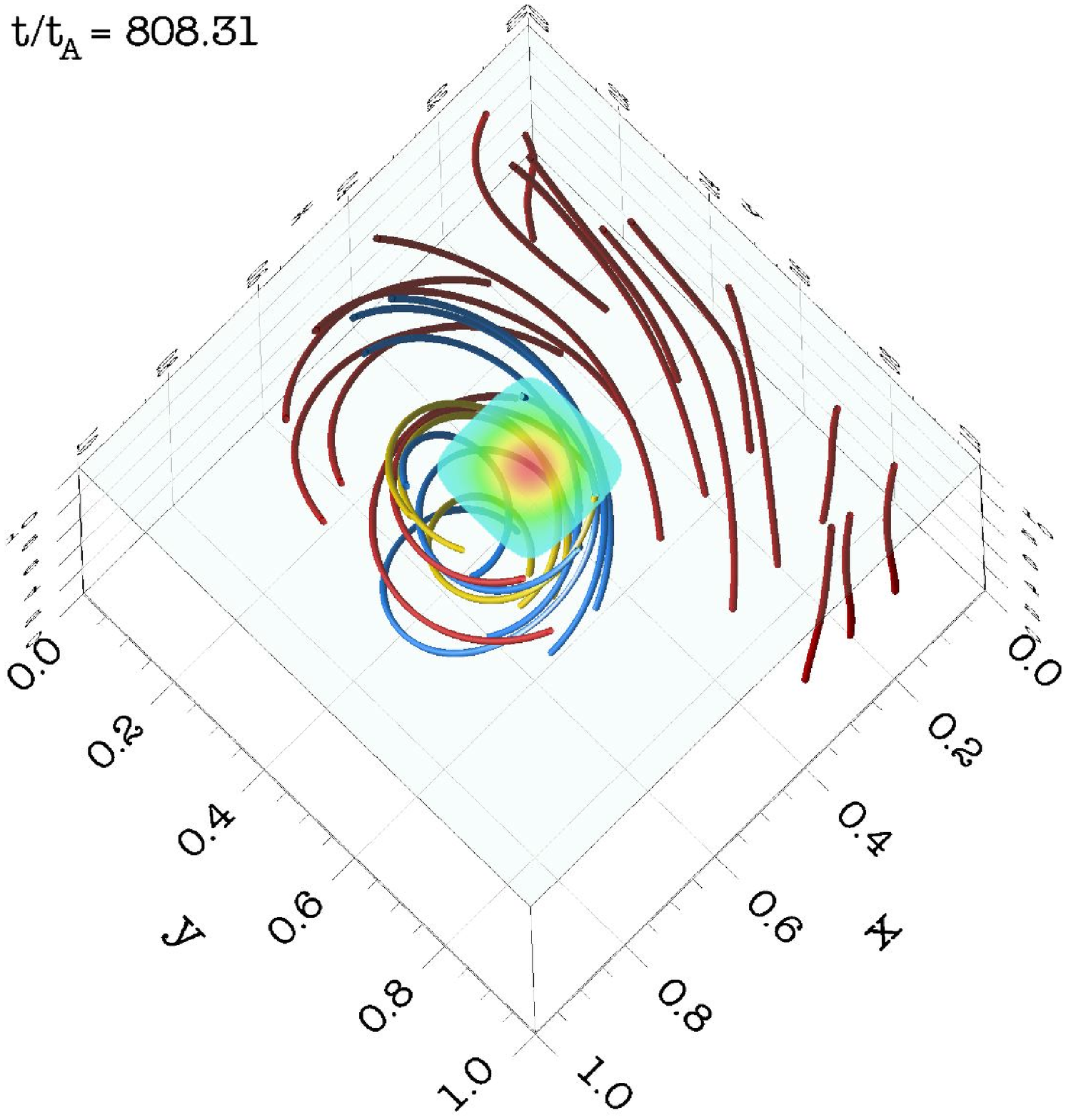}\hspace{1em}
\includegraphics[scale=.35]{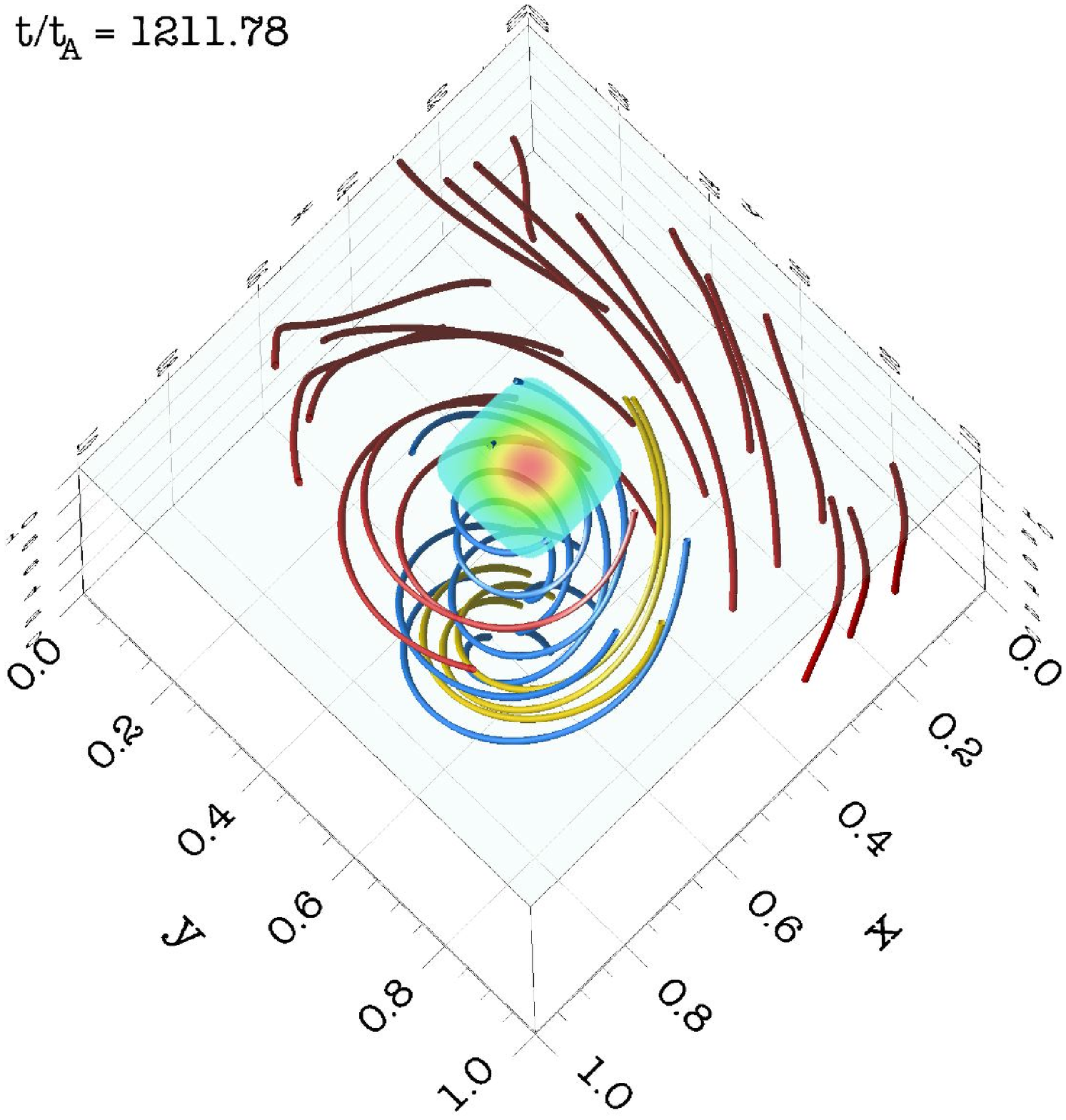}\\[.3em]
\caption{\emph{Run A}: Lateral and top views of magnetic field lines 
at selected times. In the linear stage ($t = 60.57\, \tau_A$) the
boundary vortex (shown in color in the plane $z=10$) twists
into an helix the magnetic field lines in the corresponding region
underneath the vortex. Those outside this region remain straight,
a sample of which is shown in red.
Kink instability releases magnetic energy and
untwists the field lines ($t = 100.78\, \tau_A$),
that in the nonlinear stage maintain an approximately constant
twist $\sim 180^{\circ}$.
But with time the region where field lines are twisted increases
its volume until it fills the whole computational box
($t \sim 1211.78\, \tau_A$).
The box has been rescaled for an improved visualization,
the axial length (along $z$) is ten times the length of the 
orthogonal cross section (along $x$-$y$). 
A lateral view is shown only in the first panel, at later
times a top view is preferred for a better visualization, 
as from the side the field lines appear overlapped to each other.
\label{fig:fig5}}
\end{centering}
\end{figure*}
\begin{figure*}
\begin{centering}
\includegraphics[scale=.35]{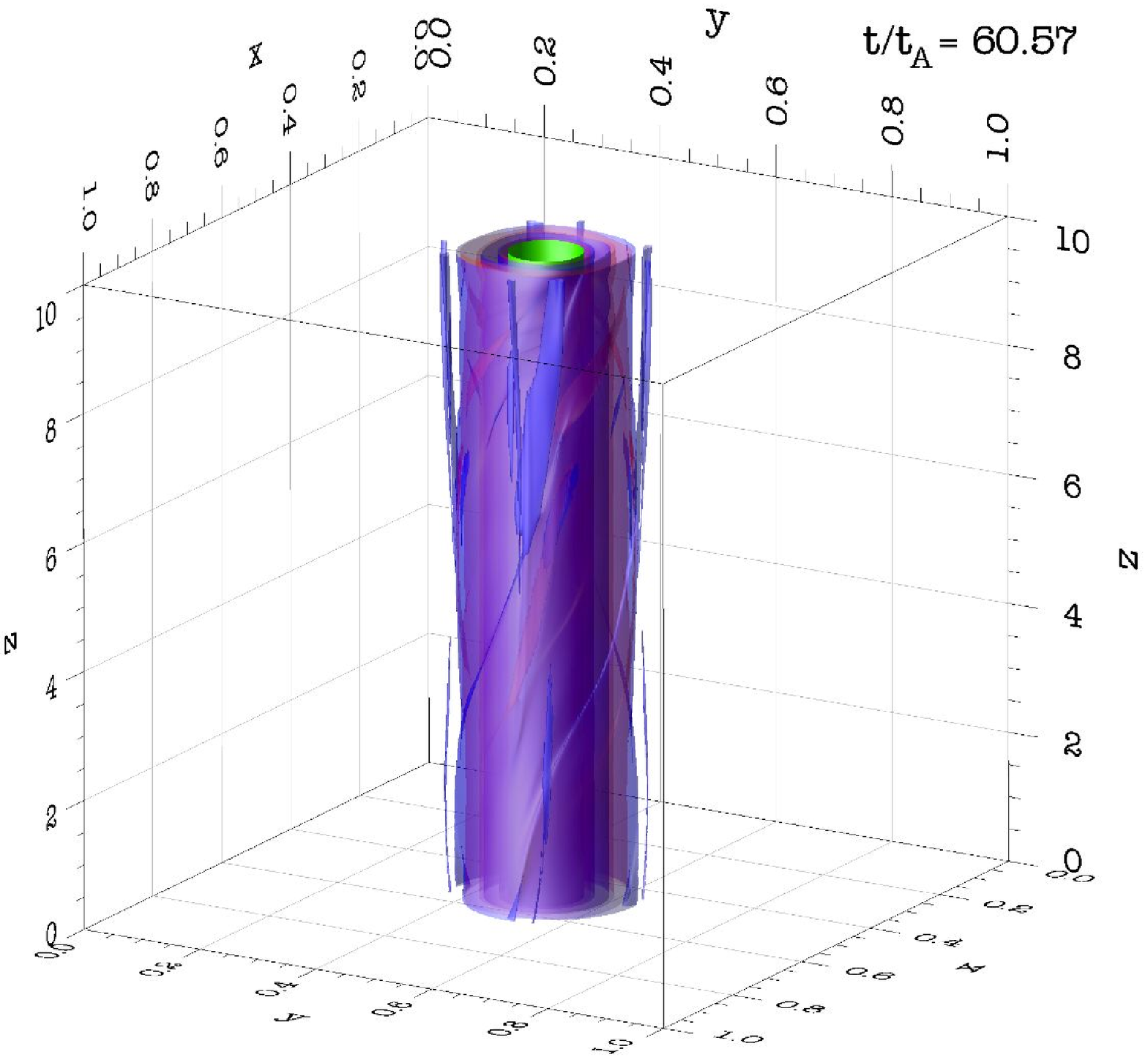}\hspace{1em}
\includegraphics[scale=.35]{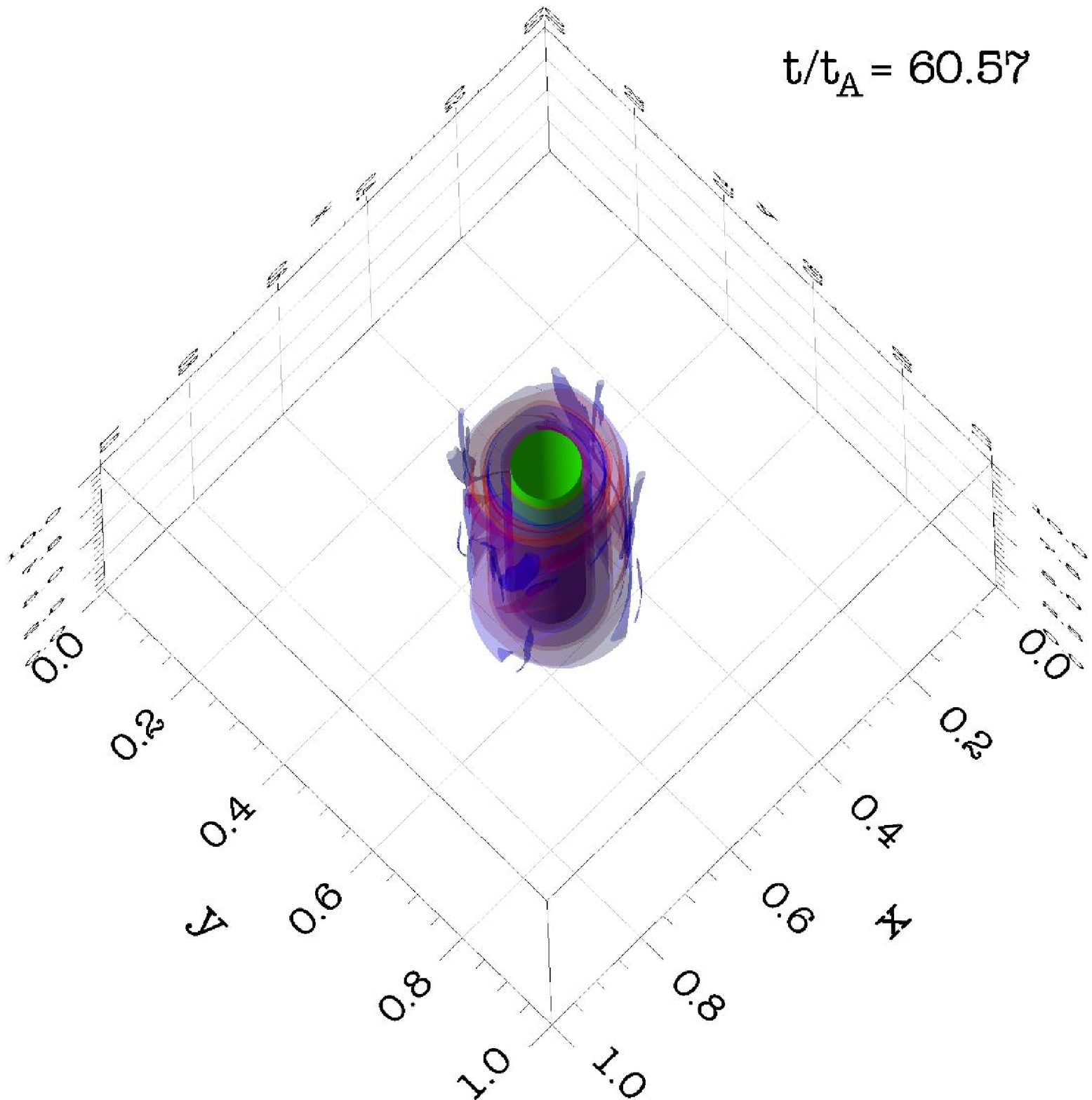}\\[.5em]
\includegraphics[scale=.35]{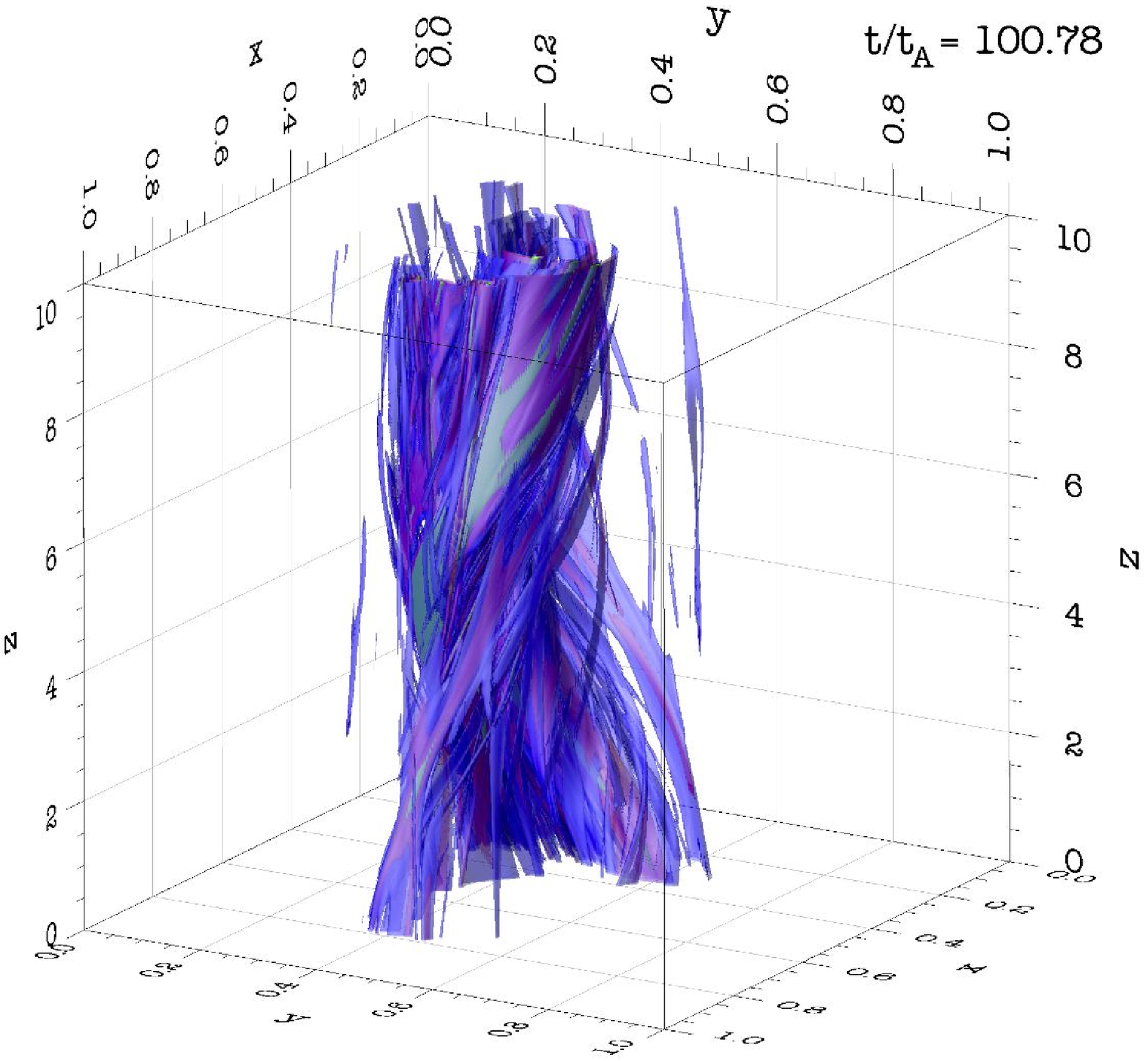}\hspace{1em}
\includegraphics[scale=.35]{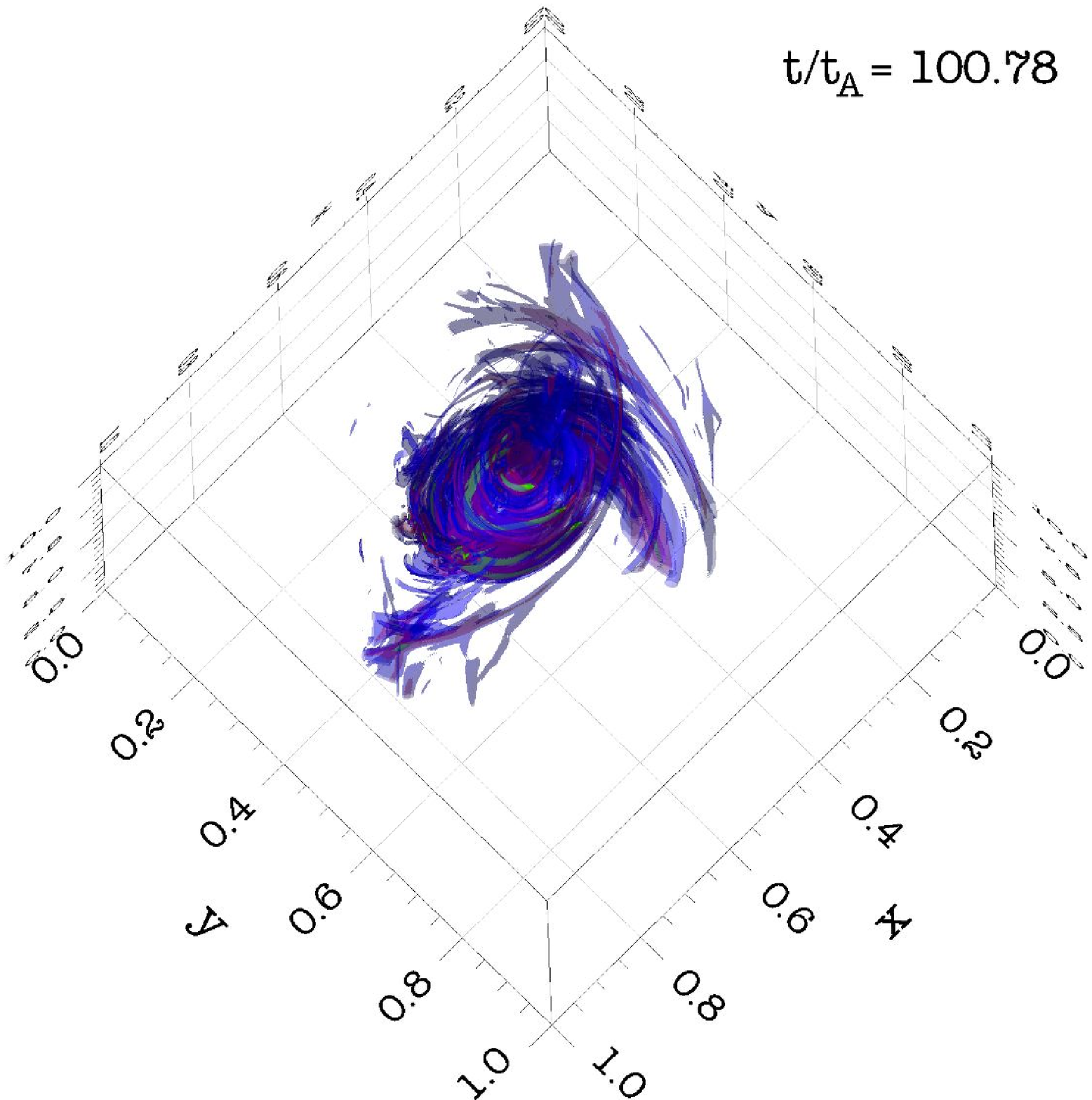}\\[.5em]
\includegraphics[scale=.35]{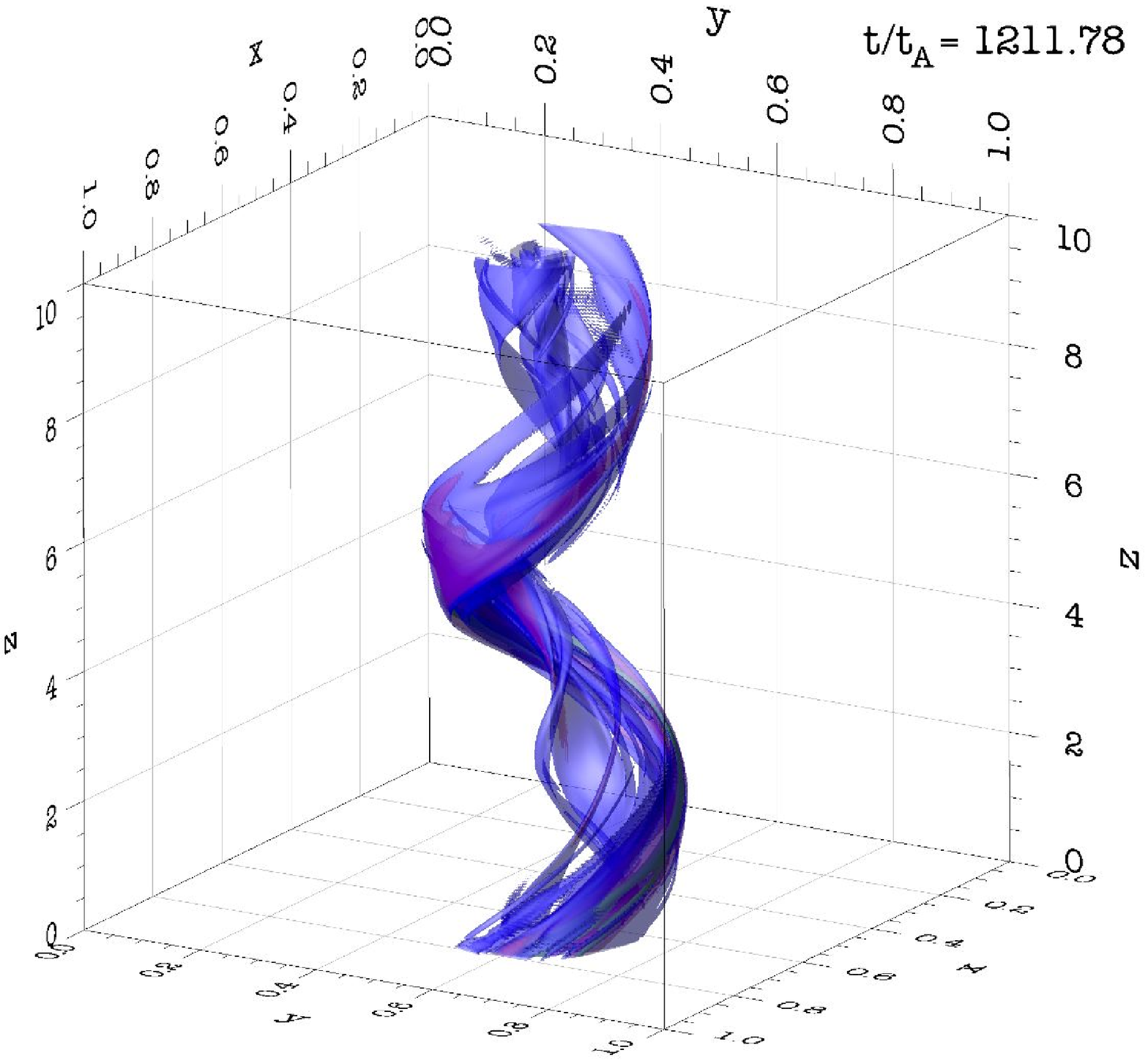}\hspace{1em}
\includegraphics[scale=.35]{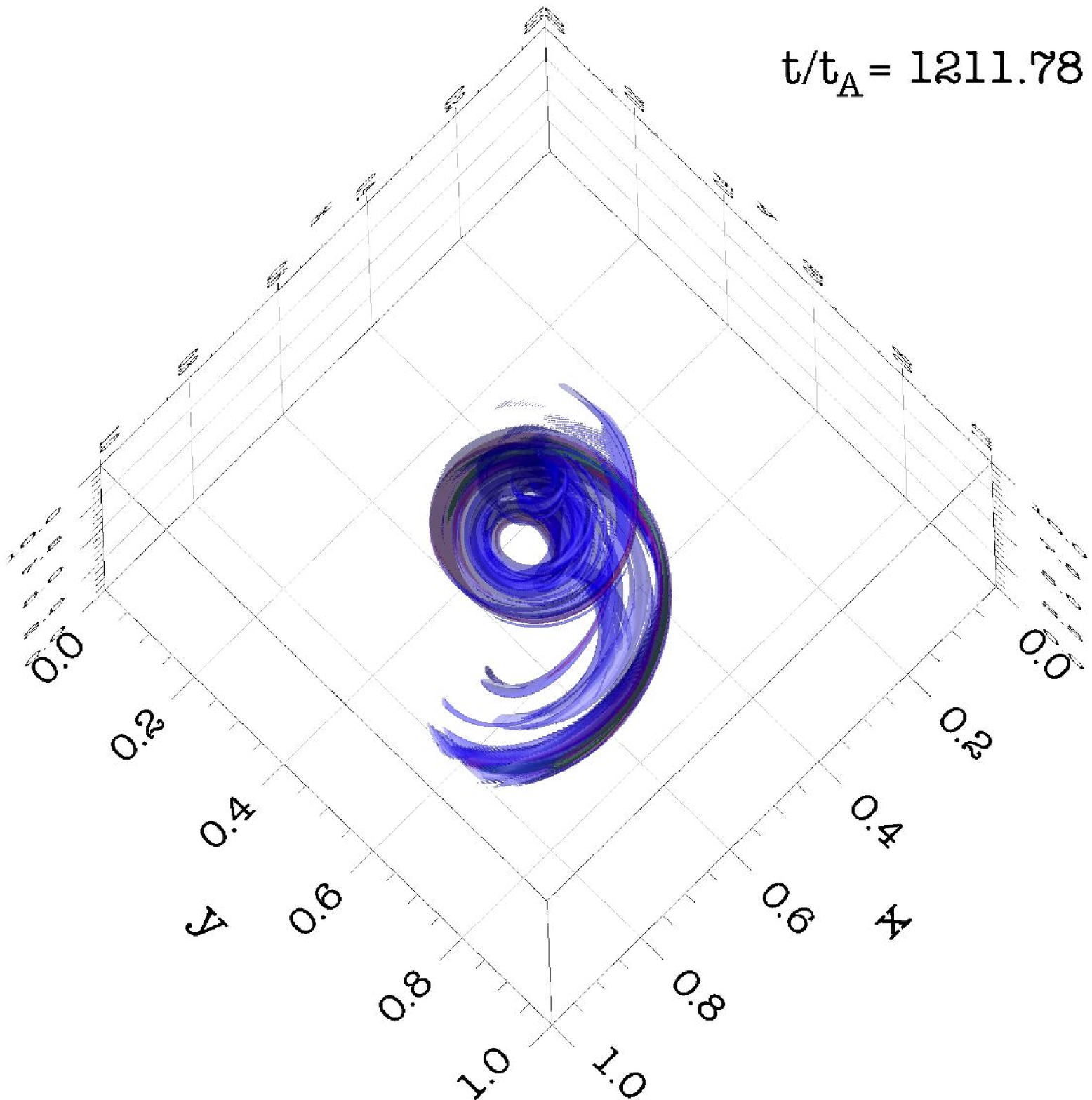}\\[.3em]
\caption{\emph{Run A}: Lateral (\emph{left column}) and top
(\emph{right column}) views of isosurfaces of the squared
current $j^2$ at selected times, respectively during
the linear stage ($t \sim 60.57\, \tau_A$), right after kink 
instability ($t \sim 100.78\, \tau_A$), and in the fully
nonlinear regime ($t \sim 1211.78\, \tau_A$).
In each panel are shown three isosurfaces of $j^2$,
corresponding respectively to $15\%$ (green), $5\%$ (red)
and $2\%$ (blue) of the maximum of $j^2$ in the box at each time.
As is typical of current sheets, isosurfaces 
corresponding to higher values of $j^2$ are nested inside those corresponding to 
lower values. 
Although the region where field lines are twisted increases
in time (Figure~\ref{fig:fig5}), the current sheets' filling
factor remains small.
The box has been rescaled for an improved visualization
as in Figure~\ref{fig:fig5}.
\label{fig:fig6}}
\end{centering}
\end{figure*}

For this reason the use of hyperdiffusion is crucial to study this problem,
otherwise diffusion dominates and a balance between the injection of energy
from the boundary and its numerical removal by diffusion is
reached very soon, inhibiting the development of kink instability
and nonlinear dynamics. This diffusive \emph{linear} regime was 
reached in previous simulations by \cite{knda09,knda10}, where 
four similar vortices were applied at the boundary. Therefore their
conclusion that nonlinear dynamics or instabilities (not to mention turbulence)
cannot develop in such physical systems is simply a numerical issue:
this can be overcome adopting hyperdiffusion as we have done here or,
alternatively, implementing grids with much higher resolutions
that require impractically large numerical resources. 

The localized boundary vortex
(shown with a colored contour in Figure~\ref{fig:fig5}) generates a mostly
poloidal magnetic field confined to the axial volume in correspondence 
of the vortex, resulting in helical field lines for the total magnetic field
(Figure~\ref{fig:fig5}, time $t = 60.57\, \tau_A$).
Outside this volume the poloidal field vanishes and only the axial field
$B_0$ is present. Amp\`ere's law then guarantees that
the total net current 
is zero. As shown in Figure~\ref{fig:fig4} in the linear stage 
($t = 0.61\, \tau_A$, $80.64\, \tau_A$) there is a stronger up-flowing
current concentrated in the middle, and a weaker ring-shaped
down-flowing current distributed at the edge of the flux-tube.

This magnetic configuration is well known to be kink unstable, and
is similar to the NC (Null Current) force-free model studied by
\cite{Lionello:1998p4478}. The main differences are that their
axial field $B_0$ is not uniform, dropping by $\sim 50\%$
outside the flux tube, and that the field lines are line-tied to
a motionless photosphere. They performed a linear stability
analysis of this configuration finding that there is a critical axial loop length 
$L_{crit}$ beyond which the system is unstable and has a constant 
growth rate $\gamma \tau_A \sim 0.02$.
They also examined other equilibria with net current finding a similar 
qualitative behavior, with variations for the critical length and growth rates.

\cite{Lionello:1998p4478} found that for the NC case 
the ratio of the axial critical length 
over the cross-length of the flux-tube is
$L_{crit}/\ell_c \sim 9$. In the case considered here the ratio of the axial
length ($L=10$) over the cross-length of the flux tube (the extent of the 
boundary vortex $\ell_c =1/4$, Equation~(\ref{eq:f0})) is $L/\ell_c = 40$, therefore
it is fully in the unstable region.
Of course at a given length (beyond the critical length) 
there is also a critical twist beyond which
the configuration is unstable. In our simulations the system is 
continuously forced at the boundary, and in the linear stage the twist
grows linearly in time (from Equation~(\ref{eq:lin1}), as the twist 
is proportional to $b_{\perp}/B_0$), thus  such a critical twist is 
certainly attained.

In our case the ``equilibrium'' solution is not static but is given by the 
linear solution (\ref{eq:lin1}),  indicated here with $\mathbf{b}_{lin}$, 
with the magnetic field growing linearly in time
while mapping the boundary vortex.  Thus we compute the perturbed
magnetic energy as
\begin{equation}
  E_{_M}^{^\star} = \int_V \mathrm{d}^{^3}\!\!x\, \left|  \mathbf{b} -\mathbf{b}_{lin} \right|^2.
\end{equation}
We find that in the linear stage this quantity grows exponentially in time, obtaining for the perturbed
magnetic field a growth rate $\gamma \tau_A \sim 0.02$, as \cite{Lionello:1998p4478}
for their NC equilibrium model. This growth rate is also confirmed by the fact
that kink instability sets in at $t \sim 83\, \tau_A$ (Figures~\ref{fig:fig2} and \ref{fig:fig3})
and $1 / \gamma \sim 50\, \tau_A$.
As mentioned in \S~\ref{par3} the forcing boundary
vortex departs from an exact circular shape at its edges where its vorticity is not exactly
constant along the streamlines, thus there is a small Lorenz force for the resulting magnetic 
field~(\ref{eq:lin1}). This small difference in the linear field acts as a perturbation.

Additionally \cite{Lionello:1998p4478} found out that configurations
with zero net current are unstable to the \emph{internal} kink mode (opposed to
the global kink mode for configurations with a net current), for which magnetic
perturbations and the radial displacement of the plasma column are confined
within the original flux tube. This is found also in our simulation as shown in
Figure~\ref{fig:fig4} at the onset of the nonlinear stage at $t = 83.85\, \tau_A$, 
when the plasma displaces inside the flux tube toward its edge where a strong 
current sheet forms.

The internal kink mode releases almost 90\% of the accumulated energy around 
time $t \sim 83.5\, \tau_A$ (Figure~\ref{fig:fig2}) in correspondence of the big
ohmic dissipative peak shown in Figure~\ref{fig:fig3}. The released energy is
$\Delta E \sim 10^3 \times 10^{22}\, erg = 10^{25}\, erg$, in the micro-flare range
(the factor to convert energy into dimensional units, given our normalization choice
discussed in \S~\ref{par3}, is $10^{22}$, i.e., $1 \rightarrow 10^{22}\, erg$).
As a result of the kink instability magnetic reconnection occurs 
(Figure~\ref{fig:fig4}, $t=85.05\, \tau_A$) and the magnetic
field lines get substantially unwind as shown in Figure~\ref{fig:fig5} (times
$t = 60.57\, \tau_A$ and $100.78\, \tau_A$) with field lines twisting only $\sim 180^{\circ}$
after the instability.

In summary, during the linear stage, the transition to and the first phase
of the nonlinear regime, the analysis of \cite{Lionello:1998p4478} is fully 
confirmed also for the photospherically driven case considered here: 
the system forced by a
circular vortex is unstable to an \emph{internal kink mode}, releases most of the stored
magnetic energy and magnetic reconnection untwists the field lines.
Linear calculations \citep{bat01} show that similar dynamics
are expected also for different configurations with different aspect ratios
and magnetic guide field values, except for those that fall
below the instability threshold.

The phenomenology described so far is also in agreement
with that of three-dimensional simulations with a realistic 
geometry \citep{al00}.
In particular strong nonlinearities persist right after 
the instability occurs ($t = 85.05\, \tau_A$ and $100.78\, \tau_A$),
when the system cannot be described as a
constant-$\alpha$ force-free state. An inverse cascade
of magnetic energy is observed, as the orthogonal
magnetic field acquires longer scales and the overall
volume occupied by twisted field lines increases,
as shown in Figure~\ref{fig:fig4} just before ($t = 85.05\, \tau_A$)
and after ($t = 100.78\, \tau_A$) the instability.
In \cite{al00} this corresponds also to an inverse cascade of magnetic
helicity, corresponding in the RMHD case to an inverse cascade
of the square potential $\psi$ (see the end of this section and our
discussion in \S~5 for more about this quasi-invariant analogous to 
helicity in RMHD).

\begin{figure}
\begin{centering}
\includegraphics[scale=.57]{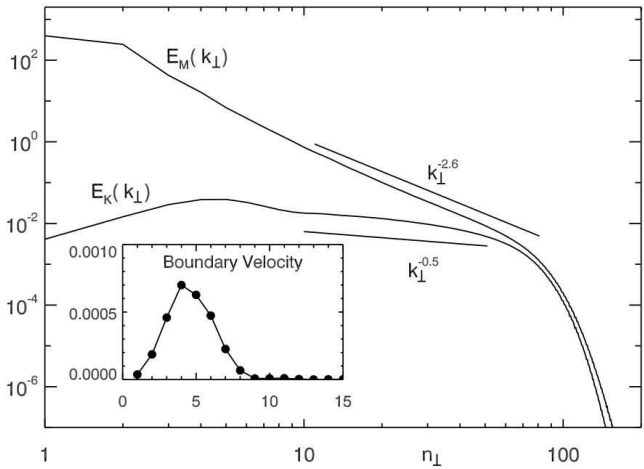}
\caption{\emph{Run A}: Magnetic ($E_M$) and kinetic ($E_K$) 
time-averaged energy spectra as a function of the orthogonal wavenumber 
$n_{_{\!\perp}}$ 
The inset shows the kinetic energy spectrum of the boundary
velocity vortex [Equations~(\ref{eq:f0})-(\ref{eq:f00})] applied at the top plate $z=10$.
\label{fig:fig7}}
\end{centering}
\end{figure}

On the other hand, at later times the dynamics are certainly
surprising when, in the fully nonlinear stage, 
fluctuations created by the kink instability are present in the corona.
For  $t > 100\, \tau_A$ magnetic
energy increases steadily, while kinetic energy remains small (Figure~\ref{fig:fig2}). 
This is in contrast
to all our previous simulations with space-filling boundary motions, either distorted
vortices \citep{rved07,rved08,rv11} or shear flows \citep{rve10}, when in the nonlinear
regime a \emph{magnetically dominated statistically steady state} was reached
where integrated quantities would fluctuate around an average value
(with velocity fluctuations  smaller than magnetic fluctuations).

In our case ohmic dissipation $J$ and the integrated Poynting flux $S$ do reach
a statistically steady state (Figure~\ref{fig:fig3}).
The integrated Pointing flux
\begin{equation} \label{eq:pflux}
S = c_A \! \int\limits_{z=L} \! \! \mathrm{d}a\, \mathbf{b}_{_\perp}  \cdot \mathbf{u}^L,
\end{equation}
is the power entering the system at the boundaries as a result of the
work done by photospheric motions on the  footpoints of magnetic field lines
($\mathbf{u}^L$ is the photospheric forcing velocity).
But in contrast to our previous results, here the power does not balance
on the average the dissipation rate, its average is slightly higher resulting
in the magnetic energy growth shown in Figure~\ref{fig:fig2}.

\begin{figure}
\begin{centering}
\includegraphics[scale=.57]{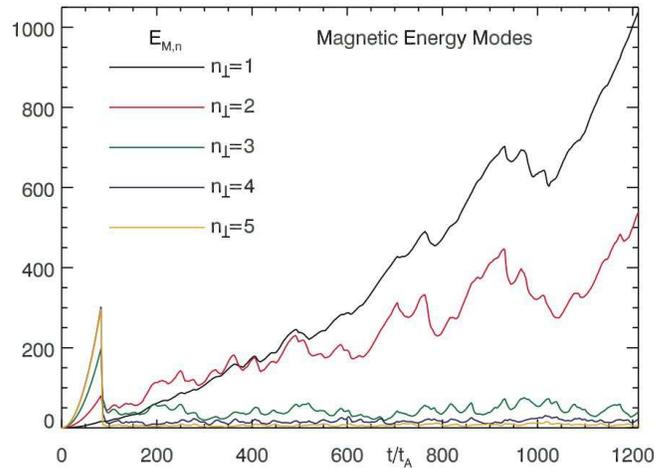}
\caption{\emph{Run A}: 
Magnetic energy modes versus time. 
While modes with wavenumber $n_{_{\!\perp}} \ge 3$
fluctuate around a mean value, the first two modes
increase steadily showing that an inverse cascade occurs. 
\label{fig:fig8}}
\end{centering}
\end{figure}

In physical space the dynamics are surprising in two ways.
\emph{First}, after the kink instability, even though we continue to stir the 
field lines' footpoints with the same vortex, no further kink instability
develops. Analogously to  the shear flow case \citep{rve10}, once the
system transitions to the nonlinear stage the  magnetic fluctuations generated
during the instability do not have a vanishing Lorentz force.
In fact around $t \sim 100\, \tau_A$, at the end of the big dissipative event,
the topology of the orthogonal component of the magnetic field is characterized
by circular, but \emph{distorted}, field lines (Figure~\ref{fig:fig4}). 
Naturally the Lorentz force does not vanish now and
the vorticity is not constant along the streamlines. 
\emph{Nonlinear terms do not vanish} as they do during the 
linear stage for $t < 83\, \tau_A$.
When they vanish  magnetic energy can be stored, without getting dissipated,
into an ordered flux-tube with helical field lines (Figure~\ref{fig:fig5}, $t=60.57\, \tau_A$),
and matching perfectly round orthogonal magnetic field lines 
(Figure~\ref{fig:fig4}, $t=80.64\, \tau_A$).
But now nonlinearity continuously \emph{transfers} energy from large to
small scales where it is dissipated.
In physical space small scales are not uniformly distributed, but they are 
organized in field-aligned current sheets. These, once
formed during the onset of the nonlinear stage, 
persist throughout the subsequent dynamics (as shown in Figures~\ref{fig:fig4} 
and \ref{fig:fig6}), with the energy cascade continuously feeding them.
\emph{Second}, the photospherical vortical motions do not
give rise to an orderly helical flux-tube as in the linear stage 
(Figure~\ref{fig:fig5}, $t = 60.57\, \tau_A$).
However, magnetic field lines get twisted, but in a disordered way 
(Figure~\ref{fig:fig5} and ~\ref{fig:fig4}, $t \geq 100.78\, \tau_A$). 
A new phenomenon occurs: on longer timescales
the magnetic field acquires longer spatial scales (Figure~\ref{fig:fig4}), 
the volume where field lines are twisted increases (Figure~\ref{fig:fig5}),
while the current exhibits always a small filling factor occupying
a small fraction of the volume (Figure~\ref{fig:fig6}).

To better understand these phenomena we need to investigate
the energy dynamics in Fourier space. We consider the spectra
in the orthogonal $x$-$y$ plane integrated along the $z$ direction.
As they are isotropic in the Fourier $k_x$-$k_y$ plane we compute
the integrated 1D spectra, so that for the total magnetic energy $E_M$
we obtain:
\begin{eqnarray}
E_M & = & \frac{1}{2} \int\limits_0^L\! \mathrm{d}z\, \iint\limits_0^{\quad \ \ell}\!
\mathrm{d}x\,\mathrm{d}y\  \mathbf{b}_{_{\!\perp}}^2 \nonumber \\
& = & \frac{1}{2} \int\limits_0^L\! \mathrm{d}z\ \ell^2 \sum_{\mathbf{k}} 
| \hat{\mathbf{b}} |^2 (\mathbf{k}, z) = \sum_{n=1}^{N} E_M(n),
\end{eqnarray}
where $n$ indicates the \emph{shell} in $k$-space with wavenumber
$\mathbf{k} = (k,l) \in \mathbb{Z}^2$ included in the range
$n-1< (k^2+l^2)^{1/2} \le n$, and $N$ is the maximum wavenumber admitted
by the numerical grid (corresponding to the smallest resolved orthogonal scale).

\begin{figure}
\begin{centering}
\includegraphics[scale=.57]{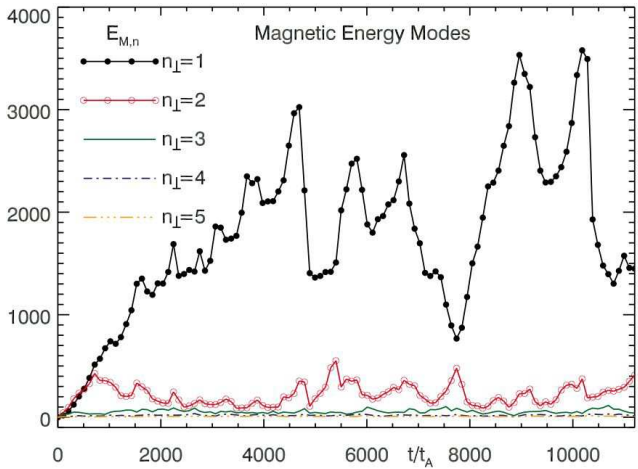}
\caption{\emph{RunB:} Over long time-scales also the
first two magnetic energy modes saturate fluctuating
around a mean value. The first mode dominates, and
the amplitude of its fluctuations corresponds to releases
of energy of $\sim 2\times 10^{25}$\,erg, in the \emph{micro-flare}
range.
\label{fig:fig9}}
\end{centering}
\end{figure}

The \emph{time averaged} magnetic and kinetic energy spectra as
a function of wavenumber are shown in Figure~\ref{fig:fig7}, the inset 
shows the spectrum of the boundary vortex' kinetic energy 
(see Equation~(\ref{eq:f0})). 
Photospheric motions therefore inject energy at wavenumbers 
between 2 and 7 (see Equation~(\ref{eq:pflux})), the system
is magnetically dominated and the power-laws exhibited at higher wavenumbers,
in the inertial range, are similar to those obtained with previous space-filling boundary
forcings \citep{rved08,rve10,rv11}, with the spectrum of magnetic energy
much steeper than that of kinetic energy.

However the time-average of the low-wavenumber modes hides an
interesting dynamics. Figure~\ref{fig:fig8} shows the first five magnetic
energy modes as a function of time. While modes with wavenumbers $n \ge 3$
after the kink instability fluctuate around a mean value, the first two modes
$n =1, 2$ grow steadily with mode $n=1$ becoming prevalent.
This shows that an \emph{inverse cascade}  takes place. While the 
\emph{direct cascade} transfers energy from the injection scale toward small scales 
(current sheets) where energy is dissipated, analogously the inverse
cascade transfers energy toward the large scales (modes 1 and 2) where 
\emph{no dissipative process} is at work and consequently \emph{energy accumulates}. 
In physical space this process gives rise to the large scales that the 
magnetic field acquires in the orthogonal direction, shown 
in Figures~\ref{fig:fig4} and \ref{fig:fig5}, discussed previously.
In the RMHD system with boundary conditions as we apply here there is no strict invariant known to follow an inverse cascade, such as magnetic helicity in 3D MHD or the square of the vector potential in 2D MHD 
\citep{bis03, ber97, bm04}.
RMHD resembles the 2D MHD case in the sense that though the square of the vector potential is not
conserved, the terms violating conservation arise only from the boundaries in the axial direction. A dynamical magnetic inverse cascade mechanism is therefore still active, impeded only by the
inputs coming from photospheric motions at the boundary, and this explains the accumulation of magnetic energy at the largest transverse scales.

\subsection{Run~B} \label{sec:runb}

The simulation described in the previous section (run~A) 
has a duration of $\sim 1200\, \tau_A$,
but this time span leaves undetermined the behavior of the 
low wavenumber modes over longer time scales.
Indeed these modes keep growing, as shown in Figure~\ref{fig:fig8}, 
resulting in a \emph{steady growth} of total magnetic energy,
shown in Figure~\ref{fig:fig2}. 
To understand the long-time dynamics
of the system, we have performed another
simulation, run~B, with the same physical parameters of run~A, but half
the orthogonal resolution (Table~\ref{tbl}), extending the duration
up to $\sim 11000\, \tau_A$. 

\begin{figure}
\begin{centering}
\includegraphics[scale=.57]{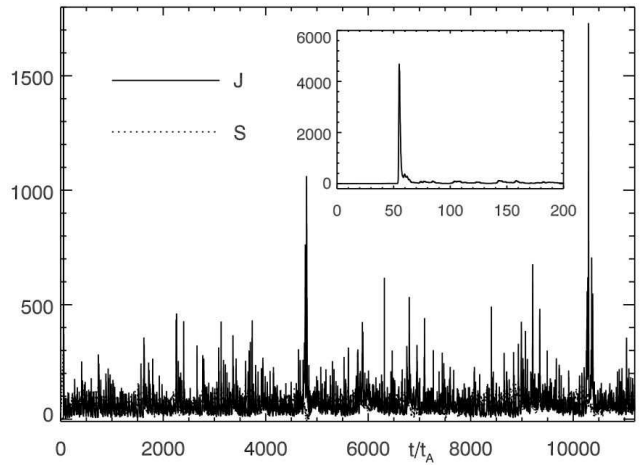}
\caption{\emph{RunB:}
Ohmic (J) dissipation rate and the integrated Poynting flux S (the injected power) versus time. Inset shows the ohmic dissipative peak corresponding to the 
development of kink instability.
\label{fig:fig10}}
\end{centering}
\end{figure}

Figure~\ref{fig:fig9} shows that over longer
times the energy of the system is prevalently in mode~1, i.e., the largest
possible scale. But this mode does not grow indefinitely and over these much
longer time-scales it reaches a statistically steady state, 
fluctuating around its mean value.
The largest energy fluctuations shown in Figure~\ref{fig:fig9} result in
energy  drops of $\sim 2000$, that in dimensional units correspond to a 
micro-flare with $\Delta E \sim 2\times 10^{25}\, erg$, releasing 
about twice the 
amount of energy released by the kink instability around 
$t \sim 85\, \tau_A$ in run~A (compare with Figure~\ref{fig:fig2}).
Notice that the kink instability does not appear in Figure~\ref{fig:fig9}
because the sampling time interval for the modes is too long in run~B,
but it is clearly shown in the r.m.s.\ of the energies (not shown) and
in the dissipation rate (see inset in Figure~\ref{fig:fig10}).

Comparing Figure~\ref{fig:fig9} with Figure~\ref{fig:fig10}, where the ohmic dissipation
rate $J$ is shown as a function of time, displays another interesting result.
The large and sharp energy drops shown in Figure~\ref{fig:fig9} correspond
to large dissipative peaks in Figure~\ref{fig:fig10}, e.g., at times $t \sim 4750\, \tau_A$
and $t \sim 10300\, \tau_A$, but it is also possible to have equally large but more
gradual energy drops, e.g., between times $t \sim 6750\, \tau_A$ and 
$t \sim 7750\, \tau_A$, without a corresponding single large dissipative peak but 
rather a cluster of smaller peaks.

\begin{figure*}
\begin{centering}
\includegraphics[scale=.57]{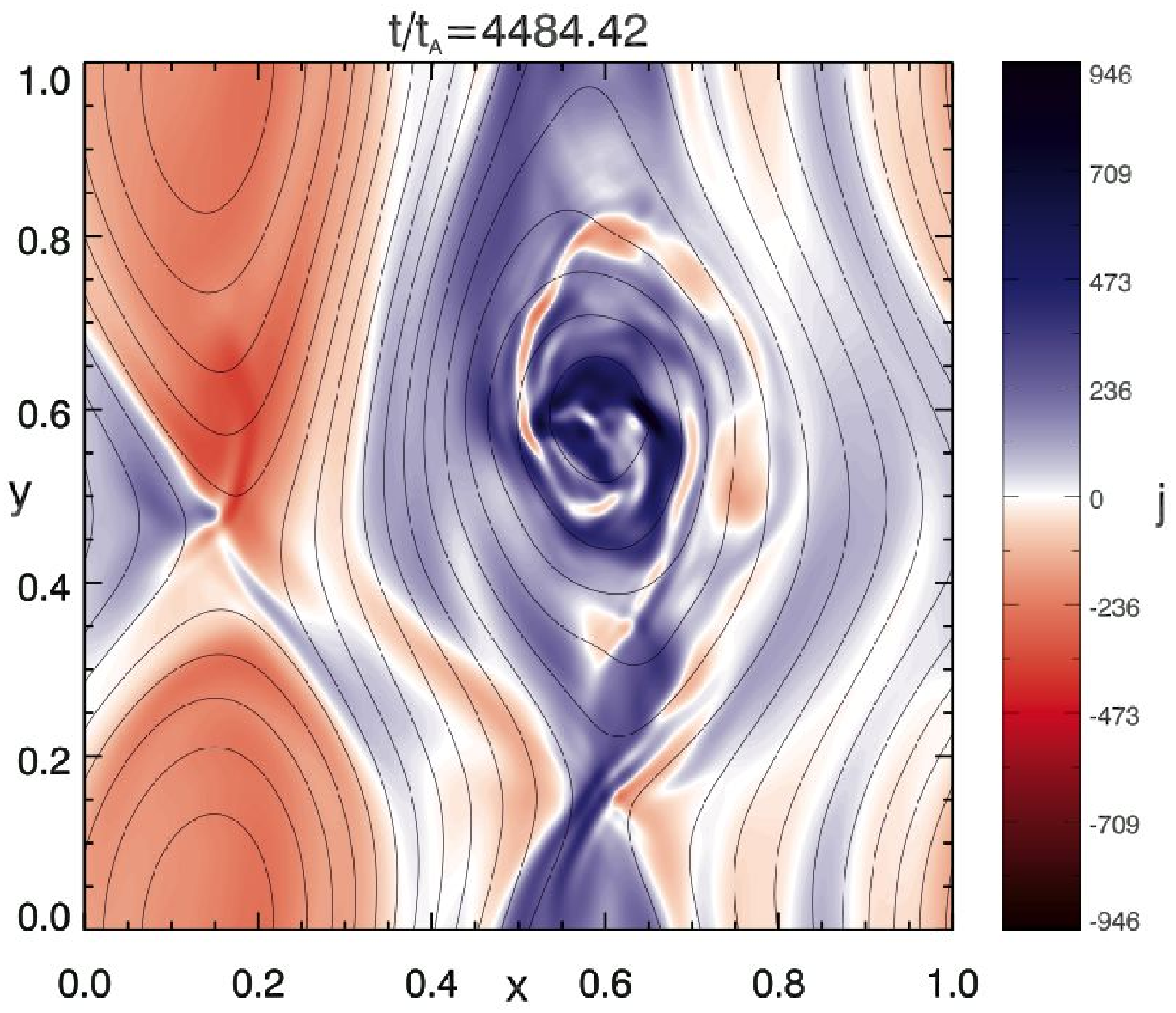}\hspace{1em}
\includegraphics[scale=.57]{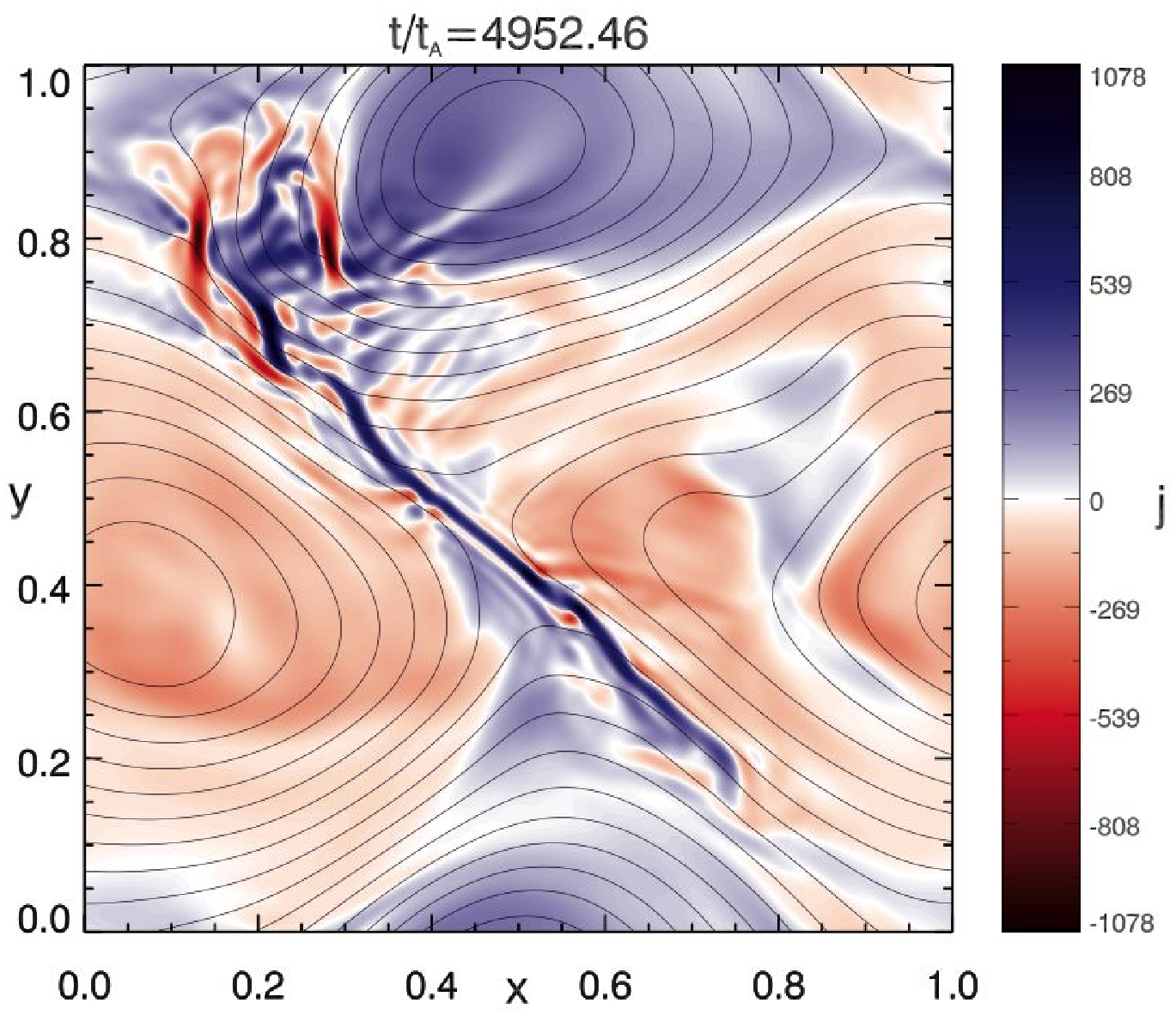}
\caption{\emph{Run~B}: Axial component of the current $j$ (in color) and field lines 
of the orthogonal magnetic field in the midplane ($z=5$) at times
$t\sim 4484\, \tau_A$ and $t\sim 4952\, \tau_A$, i.e., just before and after
the dissipative events at time $t\sim 4750\, \tau_A$ (Figures~\ref{fig:fig9}, 
and \ref{fig:fig10}).
Once the orthogonal magnetic field fills the computational box
the system oscillates between the different configurations 
with most of the magnetic energy at the large scales, through
episodes of magnetic reconnection.
\label{fig:fig11}}
\end{centering}
\end{figure*}

In physical space we have already seen in run~A that initially the inverse cascade 
corresponds to a perturbed magnetic field that occupies an increasingly larger volume
(Figure~\ref{fig:fig4}) until all the field lines in the box get twisted (Figure~\ref{fig:fig5}).
In run~B we observe that successively, once the computational box has been filled
with perpendicular magnetic field, the rising amplitude of modes ~1 corresponds to an increase
of the magnetic field intensity, while the fluctuations in the energy mode are due
to magnetic reconnection events. In fact due to the periodic boundary conditions
in $x$ and $y$ the same system repeats indefinitely along these directions.
When the orthogonal magnetic field reaches the boundary it starts to 
interact with the neighboring structures (i.e. with itself coming from the other side). The magnetic energy drops in 
mode~1 correspond to magnetic reconnection events that make the system
oscillate between the different possible configurations with energy
contained at the (large) scales of mode~1 shown in Figure~\ref{fig:fig11}
(there is no preferred orthogonal direction for the system at this scale).

While the periodic boundary conditions limit the interactions of large-scale twisted magnetic
structures it is clearly shown that interaction with such other magnetic copies of itself  is one of the ways in which
the accumulated energy can be released. Further possibilities and the 
dynamics of these interactions will be the subject of future works.

\section{Conclusions and Discussion} \label{sec:con}

In this paper we have investigated the dynamics of a closed 
coronal region driven at its boundary by a localized photospheric vortex.
Such small vortical motions with scales typical of photospheric
convection ($\sim 1000$~km) have been recently observed
in the photosphere \citep{bsg88, bms08, bms10}, and
can induce relevant dynamics in the solar corona 
\citep{vm99, wed12, pmv13}.

A ``straightened out'' closed region of the solar corona is modeled
as an elongated Cartesian box where the top and
bottom plates mimic the photosphere, and  the dynamics are integrated
with the Reduced MHD equations \citep{kp74,str76}, well suited 
for a plasma threaded by a strong axial magnetic field.

The initial condition consists simply of a uniform axial
magnetic field. Its field lines are originally straight and
its footpoints are line-tied at both ends in the top and
bottom photospheric plates.
The photospheric vortex drags the field lines' footpoints
twisting the magnetic field lines (Figure~\ref{fig:fig5}, $t=60.57\, \tau_A$). 
Even though the vortex
that we employ is not perfectly circular (Figure~\ref{fig:fig1},
Equations~(\ref{eq:f0})-(\ref{eq:f00})) in the linear stage
the field lines' tension straightens out in a round shape 
the orthogonal magnetic field lines 
(Figure~\ref{fig:fig4}, $t=0.61\, \tau_A$ and $t=80.64\, \tau_A$),
in this way the Lorentz force vanishes in the planes and the system
is able to accumulate energy.
The small departure from a round shape (at its edge) of the 
boundary vortex introduces a small perturbation in the coronal
field. The system is then unstable to the internal kink mode
(Figure~\ref{fig:fig4}, $t=80.64\, \tau_A$), and releases
about $90\%$ of the accumulated energy in a 
dissipative event (Figures~\ref{fig:fig2} and \ref{fig:fig3}).
The energy released in this event is of the order of a
\emph{micro-flare} with $\Delta E \sim 10^{25}$~erg.

These results are in agreement with those of \cite{Lionello:1998p4478},
that consider similar initial conditions, performs a refined linear
analysis, but does not employ a boundary forcing, i.e., 
the field lines are line-tied to a motionless
photosphere. Therefore the initial \emph{linear stage} and the development
of the kink instability are in agreement with previous works that
have always employed large-scale smooth fields with no 
broad-band fluctuations as initial conditions, both in the case of field lines line-tied
to a motionless photosphere \citep{Baty:1996p5040, Velli:1997p4498, 
Lionello:1998p4478, Browning:2008p4608, Hood:2009p4102}
and with a boundary driver \citep{msvh90, Gerrard:2002p5041}.

On the other hand in the solar corona perturbations are continuously injected from
the lower atmospheric layers. Numerical simulations \citep[e.g.,][]{rved08} confirm
that especially in  closed regions, where waves cannot escape
toward the interplanetary medium, broad-band magnetic fluctuations 
of the order of a few percent of the strong axial magnetic field 
(not infinitesimal perturbations as classically used in instabilities studies)
are naturally present.

Therefore in order to gain a first insight of the coronal dynamics 
when the  magnetic field is already structured, i.e., there are 
finite magnetic fluctuations (small but not infinitesimal) with small scales 
and current sheets, we continue the simulation after kink instability develops.
In fact right after kink instability the magnetic energy is small (Figure~\ref{fig:fig2}),
with $b_{\perp}/B_0 \sim 5\%$, but it is already structured with current
sheets (Figure~\ref{fig:fig4}, $t=100.78\, \tau_A$) and a broad band spectrum
(Figure~\ref{fig:fig7}).

The boundary vortex continues to twist the magnetic field lines, but
in a disordered way (Figure~\ref{fig:fig5}, $t=202.15$ -- $1211.78\, \tau_A$).
The presence of an already structured magnetic field allows nonlinear dynamics
to develop: once current sheets and small scales are present, an energy
cascade continues to feed them, as shown by the energy spectra in 
Figure~\ref{fig:fig7}. Therefore current sheets do not disappear, 
and the continuous transfer of energy from the large to the small scales prevents the field
lines to increase their twist beyond $\sim 180^{\circ}$. 
The twist remains approximately constant in the nonlinear stage as shown 
in Figure~\ref{fig:fig5} ($t=202.15$ -- $1211.78\, \tau_A$).
Furthermore because the current is now concentrated in thin 
current sheets (Figures~\ref{fig:fig4} and \ref{fig:fig6}) 
kink instabilities do not develop.

We had already observed a similar behavior in our previous simulations
that employed space-filling boundary drivers. 
In particular when the field lines were \emph{sheared} by a 
1D boundary forcing \citep{rve10} the coronal field was sheared
only in the linear stage, but after that a multiple tearing instability
developed and in the coronal field magnetic fluctuations and 
current sheets were formed, the continuous shearing motions 
at the boundary were not able to recreate a sheared coronal 
field and further instabilities were not observed.

But in the simulations presented in this paper a new phenomenon
occurs. Although the field lines' twist is approximately constant
in the nonlinear stage, the volume where field lines are twisted
increases, and the magnetic field acquires larger scales
(Figures~\ref{fig:fig4} and \ref{fig:fig5}). Besides a \emph{direct}
cascade that transfers energy from the large to the small
scales where it is dissipated in current sheets, an \emph{inverse}
cascade takes place, transferring energy from the injection
scale toward larger scales, where no dissipation takes place
and energy can accumulate. The analysis of the magnetic energy
modes (Figures~\ref{fig:fig8} and \ref{fig:fig9}) shows indeed
that on long time-scales most of the energy is stored at the largest
possible scale (mode~1). The inverse cascade is able to 
store a significant amount of magnetic energy.

Although magnetic helicity is not defined in RMHD,
the integral of the magnetic square potential $\psi$ is 
approximately conserved
(see discussion in last paragraph of \S\ref{sec:runa}).
The inverse cascade of magnetic energy corresponds
also to an inverse cascade of the square potential,
as clearly shown in Figure~\ref{fig:fig4}
where the field lines are the contour of $\psi$ (and $\psi \ge 0$).
In future compressible simulations we expect to observe
for magnetic helicity (well defined in 3D MHD) dynamics
similar to those shown here for magnetic energy, i.e., 
an increase of magnetic helicity (injected from the boundary)
and its inverse cascade, in analogy to the the inverse
helicity cascade observed by \cite{al00}.

Because of the periodic boundary conditions along $x$ and $y$ 
the system is virtually repeated along these directions.
When the field lines get twisted in the entire computational box, 
this twisted structure interacts with these neighboring twisted structures. 
This interaction is the only condition
that limits the growth of magnetic energy,
giving rise to impulsive magnetic reconnection events,
that now is not inhibited by the circular topology
of the orthogonal magnetic field lines of a single structure.
These  events make the system oscillate between the many
possible configurations with energy in mode~1 (two of these
are shown in Figure~\ref{fig:fig11}).
The associated energy drops shown in Figure~\ref{fig:fig9}
are also in the \emph{micro-flare} range with 
$\Delta E \sim 2\times 10^{25}$~erg, twice the value of the energy released
initially by the kink instability.

Although in the presented simulations the generated magnetic
structures interact only with similar structures repeated by
the periodic boundary conditions along $x$ and $y$, we can
infer that the interaction of a single twisted magnetic structure
with other magnetic structures can give rise to similar release
of energy. A more general investigations of the interaction 
between twisted magnetic structures is under way to understand
under which conditions the interaction leads to energy storage and/or
release, and to determine quantitatively these properties.

Previous simulations that employed a space-filling photospheric
forcing \citep{rved08,rve10} were not able to accumulate 
a significant amount of energy to be successively released
in micro or larger flares. Those photospheric motions, that
mimic a uniform and homogeneous convection, give instead
rise to a basal background coronal heating rate in the lower
range of the observational constraint ($10^6$~erg\,cm$^{-2}$\,s$^{-1}$)
and a million degree corona \citep{derv12}.
In the case of a space-filling boundary driver we had also
observed that the inverse cascade is inhibited  
\citep[see][\S5.4]{rved08} for typically strong DC magnetic fields. 
An inverse cascade is possible only for weak guide fields
\citep[see][\S5.4]{rved08}, a condition applicable only to 
limited regions of the corona.

We conclude that in presence of line-tying 
and a strong guide field, inverse cascade can be a good
mechanism to store energy, but only if the boundary motion is 
localized in space as the vortex used here, and not space-filling.
Subsequently the interaction of this magnetic structures with
others can release the accumulated energy.

In general photospheric motions will be a superposition
of approximately homogenous space-filling convective
motions and localized vortical and also shearing motions
\cite[e.g., see][for a localized shear case]{dlkn09}.
While the space-filling motions give rise to a basal
background coronal heating \cite[e.g.,][]{rved08}, localized motions can
give rise to higher impulsive releases of energy in the
micro-flare range and above,  contributing to coronal heating
while increasing the temporal intermittency of the energy deposition
and of its associated radiative emissions.
In future works we will consider cases with localized motions 
superimposed to a homogenous space-filling convection-mimicking
velocity field to determine, among other things, how stronger
the localized velocity has to be respect to the background motions
in order to develop dynamics similar to those presented in this paper.

As mentioned in the introduction, highly (and orderly) twisted
magnetic structures, such as \emph{flux ropes}, are used
to initiate solar eruptions (e.g., see \citet{tor11} for a recent application,
and the reviews by \citet{low01,chen11} for further examples of
this model). Kink-like instabilities developing in these flux-ropes 
give rise to an explosive dynamics leading to the formation
of a CME. We have shown that kink-unstable flux ropes are
not formed in the corona by boundary vortical motions,
unless a very strong vortex is applied and the coronal
magnetic fluctuations can then be neglected.
Therefore, although flux ropes can be formed
in the complex dynamics in and around a prominence region
\citep{alm99}, given the ubiquitous presence of magnetic
fluctuations in the solar corona, the development of 
kink-like instabilities may be strongly limited.
While the dynamics of the induced CME can be a good
approximation, we conclude that such models offer
a poor model of the initiation process for which
more realistic models are called for \citep{aal11}.

Generally speaking, in a realistic 3D geometry one might expect that the growth of energy in the transverse field
leads to an inflation and rise of a magnetic loop due to the curvature, which we have neglected here.
This effect was included by \cite{alat96},
who showed that twisting the footpoints
of a curved flux rope leads to its gradual
expansion and the system rises to larger 
solar radii. In our simulations the twist
does not increase (the overall field lines twist is limited to $180^{\circ}$),
remaining roughly constant in the nonlinear
stage. It is left to future work to understand
under which conditions such a system, including
curvature, has dynamics similar to those of
\cite{alat96}, or whether different dynamics are possible
\citep[see also][]{Gerrard:2004p79}, 
and how the dynamics develop in a 3D geometry
considering small or large-scales photospheric vortices.

\acknowledgments

This work was carried out in part at the Jet Propulsion Laboratory
under a contract with NASA. 
This research supported in part by the NASA Heliophysics Theory program NNX11AJ44G, 
and by the NSF Solar Terrestrial and SHINE programs (AGS-1063439
\& AGS-1156094), by the NASA MMS and Solar probe Plus Projects.
Simulations have been performed through the NASA Advanced 
Supercomputing SMD awards 11-2331 and 12-3188.

\end{document}